\documentclass[aps,notitlepage,nopacsnumbers,nofootinbib,english,showpacs,10pt]{revtex4}
\usepackage{amsmath}
\usepackage[dvips]{graphicx}
\usepackage[english]{babel}
\usepackage{wasysym}
\begin{document}

\title{Interaction in the dark sector}
\author{Sergio del Campo\footnote{Deceased}}
\affiliation{Instituto de F\'{\i}sica, Pontificia Universidad
Cat\'{o}lica de Valpara\'{\i}so, Av. Universidad 330, Campus
Curauma, Valpara\'{\i}so, Chile}
\author{Ram\'{o}n Herrera\footnote{E-mail: ramon.herrera@ucv.cl}}
\affiliation{Instituto de F\'{\i}sica, Pontificia Universidad
Cat\'{o}lica de Valpara\'{\i}so, Av. Universidad 330, Campus
Curauma, Valpara\'{\i}so, Chile}
\author{Diego Pav\'{o}n\footnote{E-mail: diego.pavon@uab.es}}
\affiliation{Departamento de F\'{\i}sica, Universidad Aut\'{o}noma
de Barcelona,\\
08193 Bellaterra (Barcelona), Spain}
\begin{abstract}
\noindent It may well happen that the two main components of the
dark sector of the Universe, dark matter and dark energy,  do not
evolve separately but interact nongravitationally with one
another. However, given our current lack of knowledge on the
microscopic nature of these two  components there is no clear
theoretical path to determine their interaction. Yet, over the
years, phenomenological interaction terms have been proposed on
mathematical simplicity and heuristic arguments. In this paper,
based on the likely evolution of the ratio between the energy
densities of these dark components, we lay down reasonable
criteria to obtain phenomenological, useful, expressions of the
said term independent of any gravity theory. We illustrate this
with different proposals which seem compatible with the known
evolution of the Universe at the background level. Likewise, we
show that two possible degeneracies with noninteracting models are
only apparent as they can be readily broken at the background
level. Further, we analyze some interaction terms that appear in
the literature.
\end{abstract}

\maketitle
PACS: 98.80.-k, 95.35.+d, 95.36.+x
\section{Introduction}\label{introduction}
\noindent Usually the present Universe is described by the
Friedmann-Robertson-Walker (FRW) metric coupled to the dominant
energy components, viz. baryons, dark matter (DM),  and dark
energy (DE). The two first constitute the main ingredients of
cosmic structure. The latter two comprise the so-called ``dark
sector" as their presence is felt only by their gravitational
interaction with the former.  While DM (alongside  the baryons) is
nearly pressureless, DE  is endowed with a high negative pressure
responsible for the present accelerated expansion. Since, thus
far, the information available on these two components is indirect
\textemdash they have never been directly detected, much less
produced in the lab \textemdash $\;$ their nature remains
unveiled. This may explain the very often minimalist assumption
that they evolve independently of each other, i.e., that they
interact with one another (and with whatever other component)
gravitationally only. Nevertheless, there is no reason {\em a
priori} why they should absolutely not interact otherwise;
further, a coupling in the dark sector is not only natural, it
seems inevitable \cite{polyakov1}-\cite{jmartin}.
\\  \

\noindent The simplest cosmological model with a high degree of
success from the observational side [the so-called lambda cold
dark matter ($\Lambda$CDM)] takes as DE component the cosmological
constant, or what is dynamically equivalent, the energy density of
the quantum vacuum though finely tuned to a tiny value. Alongside
this tuning problem it also presents the so-called ``coincidence
problem", i.e., the fact that nowadays the energy density of DM is
of the same order as that of the cosmological constant despite the
former declines with cosmic expansion as $a^{-3}$ (where $a$ is
the scale factor of the FRW metric) while the latter remains fixed
\cite{coincidence}.
\\   \

\noindent A way to alleviate, or even solve, the latter problem is
to allow for some nongravitational interaction in the dark sector
that gives rise to a continuous transfer of energy from DE to DM.
(Note that a transfer in the opposite direction would only
exacerbate the problem, so it is more appealing to assume that the
transfer occurs just in that sense; although the opposite
possibility is not excluded, neither is the possibility of one or
more changes in the direction of the transfer as the Universe
expands). Current data are compatible with such transfer although
the evidence is, so far, not fully conclusive
\cite{prd_german,prd_bin-elcio,abdalla-ferreira}. For reviews on
the subject see \cite{bolotin,fernando} and references therein.
See, also, the recent contributions \cite{copeland,langlois}.
\\  \

\noindent Given our ignorance about the nature of the components
of the dark sector (to begin with, we do not even know whether
these are single or multiple components), there is no reliable
guidance to determine the interaction term. Therefore, the
expressions proposed in the literature (except for the obvious
restriction that such term must be small, not to deviate seriously
from the predictions of the $\Lambda$CDM model) are necessarily
phenomenological and based on heuristic arguments. Here we propose
a different route to arrive at reasonable expressions of such term
based on the likely evolution of the ratio between the densities
of DM and DE.
\\  \

\noindent From the history of structure formation we know that the
former dominated at early times (though preceded by a short period
of radiation prevalence), and from the current accelerated
expansion we infer that the latter started to dominate in the
recent past, and it will likely dominated for ever \textemdash at
least if the $\Lambda$CDM or a slight variant of it turns to be
the right cosmological model. Therefore, qualitatively speaking,
the said ratio must decrease monotonously  for most of the time.
Although, as argued below, at very high redshifts it must have
varied little or even stayed constant, and it will also scarcely
vary  or be constant in the far future. Accordingly, by wisely
parametrizing the said ratio and using the conservation equations
of the energy components, valid in most theories of gravity,
compelling expressions for the interaction term are readily
obtained.
\\  \

\noindent This paper is organized as follows. Section II relates
the interaction term with the said ratio, establishes the criteria
that any reasonable parametrization of the latter should comply
with and illustrates it by proposing three different
parametrizations  based solely on the FRW metric and the energy
conservation equations. Section III makes use of recent data on
the Hubble history to constrain the free parameters entering these
parametrizations. Section IV deals with two apparent degeneracies
between the interaction term and the equation of state (EOS) of
DE. Section V considers different proposals for the interaction
term made in the literature  to see whether their corresponding
ratios between the energy densities comply with the criteria  of
SEc. II. Finally, Sec. VI summarizes our findings.
\\  \

\noindent As usual, a subscript zero attached to any quantity
means that the latter is to be evaluated at present time.

\section{Parametrizing the ratio DM/DE}
\noindent  Let us consider a homogeneous and isotropic universe
whose main energy components are baryons, dark matter  and dark
energy, subscripts $b$, $m$, and $x$, respectively. The first two
being pressureless, the latter with a not necessarily constant EOS
parameter $w = p_{x}/\rho_{x} < - 1/3$.
\\  \

\noindent We first look for an equation describing the dependence
of the quantity, $Q$, that gauges the strength of the interaction
in the dark sector, on the ratio between the densities of DM and
DE, $r \equiv \rho_{m}/\rho_{x}$. Thus, we write the conservation
equations of the components as
\begin{eqnarray}
\dot{\rho}_{b} & + & 3H \rho_{b} = 0 \, , \\
\dot{\rho}_{m} & + & 3H \rho_{m} = Q \, ,  \\
\dot{\rho}_{x} & + & 3H (1+w) \rho_{x} = -Q \, ,
\label{eq:conservation}
\end{eqnarray}
where $H \equiv \dot{a}/a$ denotes the Hubble function. In writing
last two equations we have implicitly assumed that all DM and all
DE participate in the interaction. For $Q > 0$ energy flows from
DE to DM and vice versa for $Q$  featuring the opposite sign;
however, as mentioned above, we will take $Q > 0$ as it alleviates
the coincidence problem. In accordance with strong constraints
from local gravity experiments \cite{peebles-ratra,hagiwara} we
assumed that the baryons conserve separately.
\\ \

\noindent As it is readily derived, the condition $Q \geq 0$
imposes $\dot{r}/r \geq 3Hw$, i.e., it restricts the rate at which
$r$ can decrease as the Universe expands. (Recall that in the
$\Lambda$CDM model ($w = -1$) one has $\dot{r} = -3Hr$).
\\  \

\noindent From the above, we get
\begin{equation}
Q = \left(\frac{a \, r'}{r} \, - \, 3 w\right)H\,
\frac{\rho_{m}}{1+r} \, , \label{eq:Q1a}
\end{equation}
where the prime stands for $d/da$.
\\  \

\noindent We reasonably expect $Q$ to be smaller than the second
term on the left-hand side of (2). This, alongside the restriction
$Q > 0$, leads to $0 < ar' - 3wr < 3(r+r^{2})$.
\\   \

\noindent We now look for reasonable expressions for $\, r \,$ to
be inserted in (\ref{eq:Q1a}); this, by passing, will illustrate
our method. Any expression for $\, r \,$ should satisfy $\, {\rm
d}r/{\rm d}a <0 \, $, except when $\, a \rightarrow 0 \, $ and
when $\, a \rightarrow \infty \,$, because in these limits $\, r
\,$ must either state constant or vary very slowly. Hence the
quantities $\, r_{+} \equiv r(a \rightarrow 0)$, $r_{0} \equiv r(a
= 1) \simeq 25/70\, $,\footnote{For simplicity, we take the
current energy budget as: baryons $5\%$, DM $25\%$, and DE
$70\%$.} and $r_{-} \equiv r(a \rightarrow \infty)\,$ must fulfill
$0 \leq r_{-} < r_{0} < 1 < r_{+}$, with $\, r_{+} \,$ finite.
Should the latter diverge, it would be because of one of the three
following possibilities: (i) $\rho_{x}$ vanished and $\rho_{m}$
stayed finite as $a \rightarrow 0$. Then, the initial amount of DE
would be zero and no transfer of DE to DM could ever occur whence
$Q$ would vanish altogether at all times. (ii) $\rho_{x}$ remained
bounded but $\rho_{m}$ diverged (in the same limit as above). In
this second case Eq. (2) implies that $Q$ would diverge, and by
(3) $\, \dot{\rho}_{x}$ would become minus infinity. This, in
turn, would lead to the instantaneous transfer of the total amount
of DE to DM. That is to say, the dark energy component would
disappear completely for good. However,  if $\rho_{m}$ would obey
$\, \rho_{m} = {\rm constant}\, a^{-3} + \epsilon(a) \, $ where
$\epsilon$  does not vanish as $\, a \rightarrow 0$, things would
be different, because in this instance, $Q \, $ does not diverge.
This introduces an extra uncoupled DM parameter that could be
included when comparing with the data. Nevertheless, the
expression for $\, \rho_{m} \, $ corresponds to the case where not
all DM interacts, only a part of the total, but this was already
excluded from our assumptions above. (iii) Both energy densities
diverge, but $\rho_{m}$ tends to a higher infinity than $\rho_{x}$
does. Clearly, this case is the same as the previous one. Finally,
the expression for $\, r \, $ must gently interpolate between
$r_{+}$ and $r_{-}$. Observational data suggest that the former is
much larger than unity;  see, e.g., \cite{prd_bean,jcap_xia-viel}.
\\  \

\noindent The possibility that for some period between $r_{+}$ and
$r_{0}$ the ratio $r$ is not decreasing would imply a severe
deviation from the $\Lambda$CDM model that, most likely, would
bring observable consequences on structure formation and the
integrated Sachs-Wolfe effect. This is why we assume that $\, {\rm
d}r/{\rm d}a < 0 \,$, except possibly at $r_{+}$ and $r_{-}$.

\subsection{First parametrization}
\noindent A possible choice for the ratio $\, \rho_{m}/\rho_{x} \,
$ is
\begin{equation}
r = r_{-} \, + \,  \frac{r_{+} - r_{-}}{1 \, + \, \beta
a^{\alpha}} \, , \label{eq:r(a)1}
\end{equation}
where $\alpha$ and $\beta$ are positive-definite constants. As it
can be readily checked (\ref{eq:r(a)1}) complies with the above
conditions, and the corresponding current ratio is $\, r_{0} =
r_{-} + (r_{+} - r_{-})(1+\beta)^{-1}$. If, as it seems reasonable
(at least for $w = {\rm constant}$), $r_{+} \gg r_{-}$, then
$r_{+} \simeq r_{0} \, (1+\beta)$. This suggests $\beta \gg 1$.
\\  \

\begin{figure}[!htb]
  \begin{center}
    \begin{tabular}{c}
      \resizebox{120mm}{!}{\includegraphics{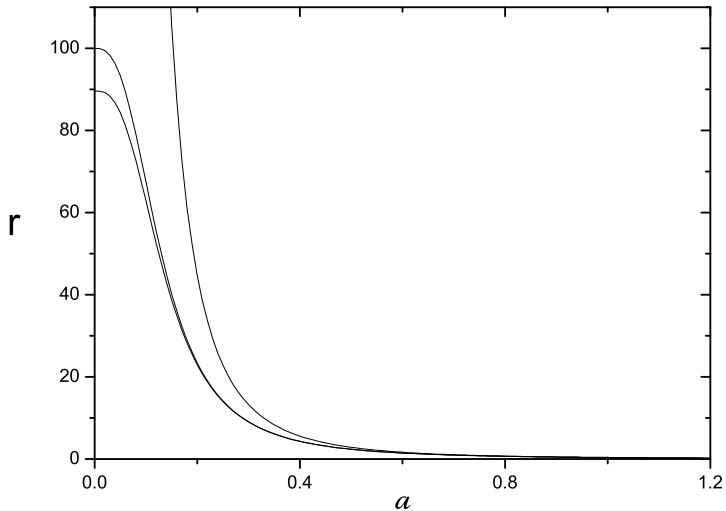}}\\
    \end{tabular}
    \caption{The  first curve (from left to right) shows the
    evolution of ratio $\rho_{m}/\rho_{x}$ corresponding to the expression (\ref{eq:r(a)1})
    where we have used $\alpha = 2.70$ and $\beta = 89.6$, consequently $r_{+}
    = 32.36$.The same can be said for the next curve but with
    $\alpha = 2.76$, $\beta = 279$, and $r_{+} = 100$. The right-most curve
    depicts for comparison, the evolution of the ratio
    for the concordance $\Lambda$CDM model, $\rho_{m}/\rho_{\Lambda} = r_{0} \, a^{-3}$.
    In all three cases, we have taken $r_{0} = 25/70$. As it is apparent, the
    interaction $Q > 0$ mitigates the coincidence problem.}
    \label{fig:r1}
 \end{center}
\end{figure}

\noindent From the condition $Q \geq 0$, it follows that $\, ar'/r
> 3w$. This alongside the observationally well-supported
assumption $\, r_{+} \gg r_{-}$,  yields
\begin{equation}
\frac{a\, r'}{r} \simeq - \frac{\alpha \beta a^{\alpha}}{1+ \beta
a^{\alpha}} \geq 3w \, ,
\label{eq:arprime}
\end{equation}
thereby in the limit $a \gg 1$,  the constraint $\alpha \leq 3
\mid w \mid \, $ follows.
\\  \
\noindent The first derivative of $\, r$, evaluated at present
time ($a =1$) and under the assumption $r_{-} \ll r_{+} $, yields
\begin{equation}
r'(a= 1) \simeq - r_{0} \frac{\beta}{1 + \beta}\, \alpha \simeq -
r_{0} \alpha.
\label{eq:rprimepresentime}
\end{equation}
Because  $0 \leq \alpha \leq 3$, it is seen  that $\mid r'(a=
1)\mid \leq 3 r_{0}\, $ whereby the slope of $r$ at $a = 1$ is
more gentle than the corresponding one of $\Lambda$CDM model,
$3r_{0}$, whence the coincidence problem gets alleviated.
\\  \
\noindent Figure \ref{fig:r1} shows the evolution of $r$ for some
chosen values of the free parameters compatible with observational
data, at least,  at the background level.

\noindent In the particular but interesting case that $w = -1$ and
$r_{+} \gg r_{-}$ we have
\begin{equation}
Q \simeq \left(3 \, - \, \frac{\alpha \beta a^{\alpha}}{1 \, + \,
\beta a^{\alpha}} \right)\, H \,  \frac{1+\beta\,
a^{\alpha}}{r_{+}+1+\beta\, a^{\alpha}} \, \rho_{m}.
\label{eq:Q1stparm}
\end{equation}
This illustrates how a reasonable expression for the interaction
term between DM and DE can be devised from simple arguments based
on the evolution of the ratio $\rho_{m}/\rho_{x}$.
\\  \

By inserting the last expression in Eq. (2) we obtain, after
integration, the evolution of the DM density,
\begin{equation}
\frac{\rho_{m}}{\rho_{m0}} = a^{-\frac{3 r_{+}}{1+r_{+}}}
\left[\frac{1+r_{+}+\beta
a^{\alpha}}{1+r_{+}+\beta}\right]^{-\frac{\alpha(1+r_{+})-3r_{+}}{\alpha(1+r_{+})}}.
\label{eq:rhom1}
\end{equation}
In the limiting cases $a \ll 1$ and $ a\rightarrow \infty$, one
has $\, \rho_{m} \propto a^{- 3r_{+}/(1+r_{+})}\, $ and $\,
\rho_{m} \propto a^{- \alpha}\, $, respectively. These results are
very reasonable. At early times the interaction has  not had much
chance to be felt; besides, the second term on the left-hand side
of (2) is much larger than $Q$, thereby the DM density goes down
practically as if it were no interaction, $\rho_{m} \sim a^{-3}$.
At late times, the interaction has had plenty of time to be felt
and $Q$ is not so small as compared with the said second term;
thus $\rho_{m}$ departs from the case of just pure expansion and
goes down more slowly.
\\  \

\begin{figure}[!htb]
  \begin{center}
    \begin{tabular}{c}
      \resizebox{120mm}{!}{\includegraphics{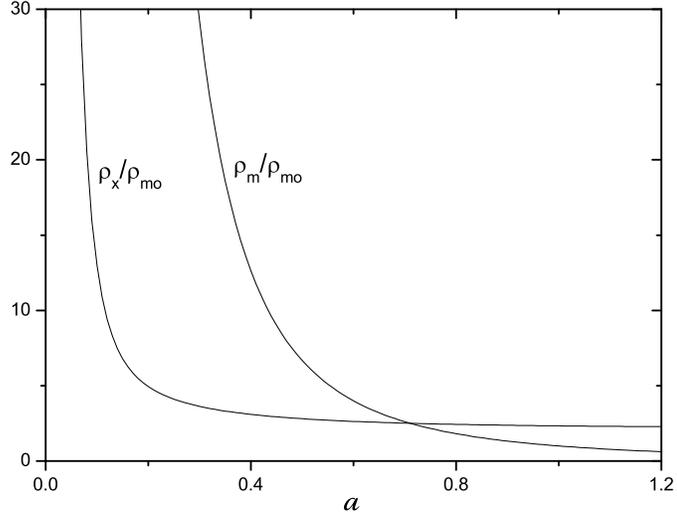}}\\
    \end{tabular}
    \caption{The curve on the right specializes  Eq. (\ref{eq:rhom1}) for
    $\alpha = 2.5$, $r_{+} = 100\, $, and $\, \beta = 697/3$. The curve on the left
    results from dividing the previous one  by $\, r$, as given by Eq. (\ref{eq:r(a)1}).}
    \label{fig:sol1}
 \end{center}
\end{figure}

\noindent Figure \ref{fig:sol1} depicts, for illustrative
purposes, the evolution of the ratios $\, \rho_{m}/\rho_{m0}$ and
$\, \rho_{x}/\rho_{m0}$ for some selected values of the free
parameters.

\subsection{Second parametrization}
\begin{figure}[!htb]
  \begin{center}
    \begin{tabular}{c}
      \resizebox{80mm}{!}{\includegraphics{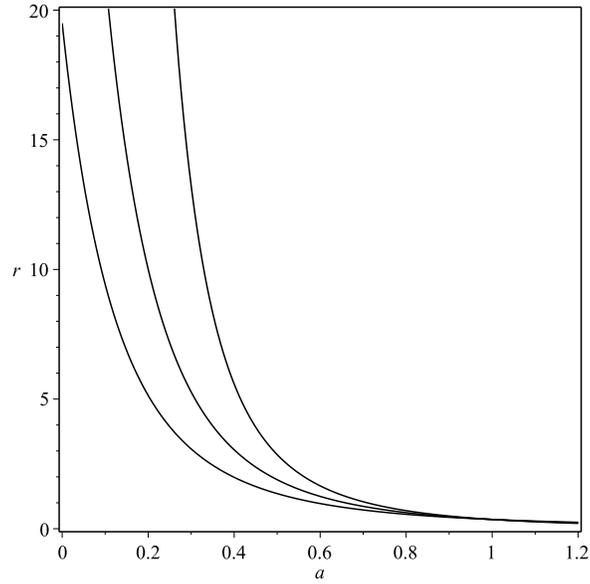}}\\
    \end{tabular}
    \caption{The first two curves from left to right, with $\lambda = 8$ and $10$,
    respectively, and $r_{0} = 25/70$,
    depict the  ratio $\rho_{m}/\rho_{x}$ described by Eq. (\ref{eq:ra(a)5}).  Though not apparent
    in the figure these ratios do not vanish asymptotically, they tend to $r_{-} =(25/70)\,e^{-4}$,
    and $r_{-} =(25/70)\,e^{-5}$, respectively. The third curve,
    shows, for comparison, the evolution  of the  ratio for the $\Lambda$CDM model,
    namely, $ \rho_{m}/\rho_{\Lambda} = r_{0} \, a^{-3}$ with an identical value for $\, r_{0}$.
    Again, the interaction is seen to alleviate the coincidence problem.}
    \label{fig:ra(a)5}
 \end{center}
\end{figure}

\noindent Another possible expression for the ratio
$\rho_{m}/\rho_{x}$ is
\begin{equation}
r= r_{-} \, \exp \left(\frac{\lambda}{1+a}\right) \, ,
\label{eq:ra(a)5}
\end{equation}
where $r_{-} = r(a \rightarrow \infty)>0$ and $\lambda>0$. On the
other hand, $r_{+} \equiv r(a \rightarrow 0) =  r_{0} \,
e^{\lambda/2}$; see Fig. \ref{fig:ra(a)5} for the cases $\lambda =
8$ and $10$, both with  $r_{0} = 25/70$.
\\  \
\noindent The condition for alleviating the coincidence problem
$\, \mid r'(a = 1)\mid < 3\, r_{0}\, $ implies $\lambda < 12$. For
$w = - 1$, this upper bound on $\lambda$ also ensures  $Q > 0$.
\\  \

\noindent Inserting (\ref{eq:ra(a)5}) in (\ref{eq:Q1a}) and
assuming $w = -1 $, we get
\begin{equation}
Q = \left(3 \, - \, \frac{\lambda a}{(1+a)^{2}}\right)\, H \,
\frac{\rho_{m}}{1\, +\, r_{-} \, e^{\lambda/(1+a)}},
\label{eq:Q2ndparm}
\end{equation}
whence the corresponding differential equation for the DM density
is
\begin{equation}
a\, \rho'_{m} \, +\, \left\{3\, + \, \left[\frac{\lambda \,
a}{(1+a)^{2}} \, - \, 3 \right]\, \frac{1}{1\, + \, r_{-} \, \,
e^{\lambda/(1+a)}} \right\} \, \rho_{m} = 0 \, .
\label{eq:rhomevol4}
\end{equation}
\\  \

\noindent This has analytical solutions in the limiting cases, $a
\rightarrow 0\, $  and $a \rightarrow \infty$; namely, $ \rho_{m}
\propto a^{-3r_{+}/(1+r_{+})} \, $ and $\, \rho_{m} \propto
a^{-3r_{-}/(1+r_{-})}$, respectively, where $\, r_{-} = r_{0} \,
e^{-\lambda/2} >0$; i.e., $r \, $ never vanishes. Notice that, as
in the previous case, the limiting solutions are rather
reasonable; for small $a$, only a tiny amount of energy can be
transferred to DM whence $\rho_{m}$ decreased practically as
though $Q$ vanished, $\rho_{m} (a \ll 1) \sim a^{-3}$. For large
$a$, a large amount of energy has already been transferred,
significantly compensating the ``natural" decrease of density;
hence, $\rho_{m} (a \gg 1) \simeq {\rm constant}$. This is
apparent in Fig. 4, which depicts the numerical solution
(\ref{eq:rhomevol4}) for the cases $\lambda=8$  (i.e.,  $r_{-}
=(25/70)\, e^{-4}$), and $\lambda=10$ [i.e.,  $r_{-} =(25/70)\,
e^{-5}$], respectively.

\begin{figure}[htb]
\hspace{-2.5cm}
\begin{minipage}{0.3\textwidth}
\centering
 \includegraphics[width=8cm]{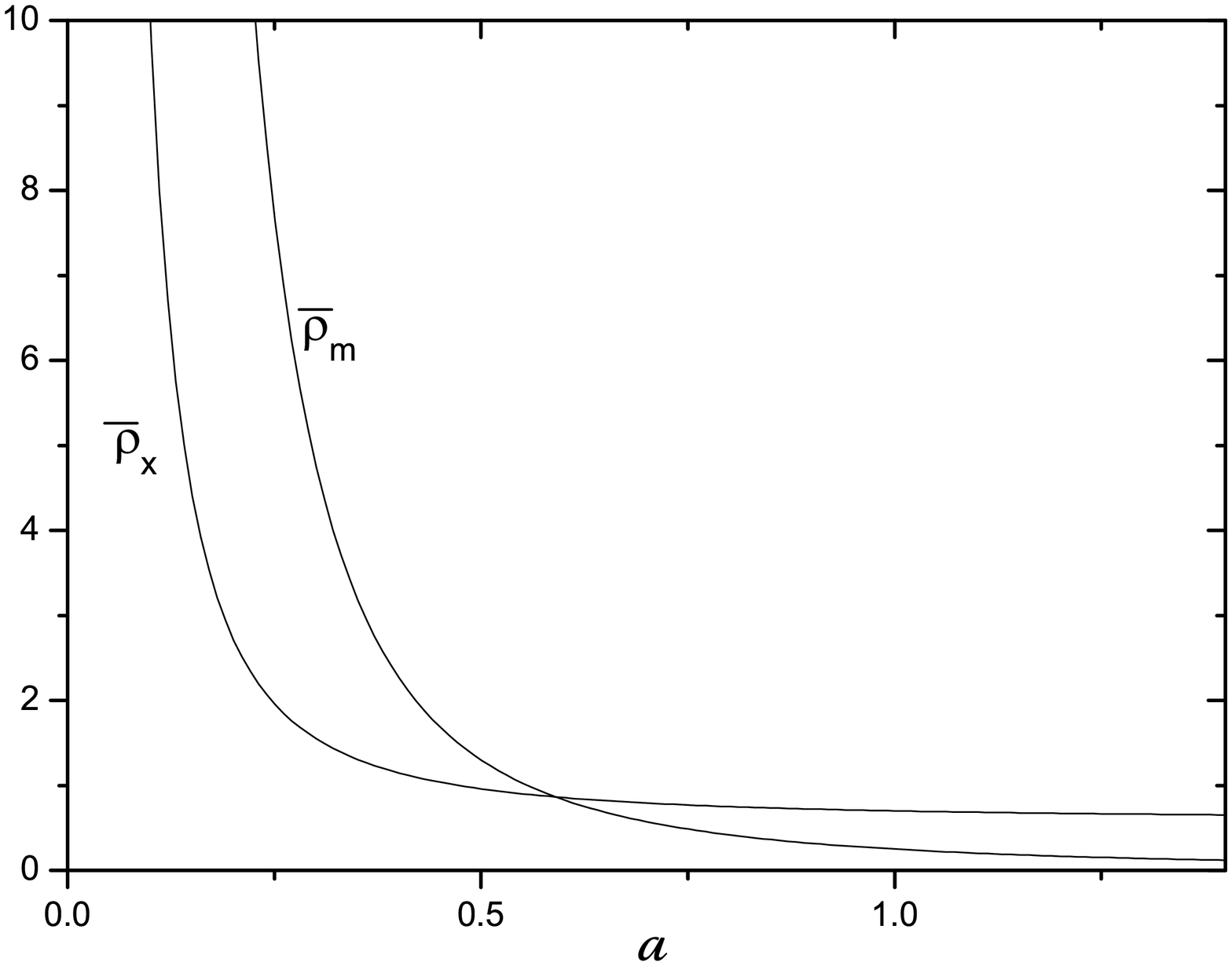}
 \end{minipage}
 \hspace{4 cm}
\begin{minipage}{0.3\textwidth}
\centering
 \includegraphics[width=8cm]{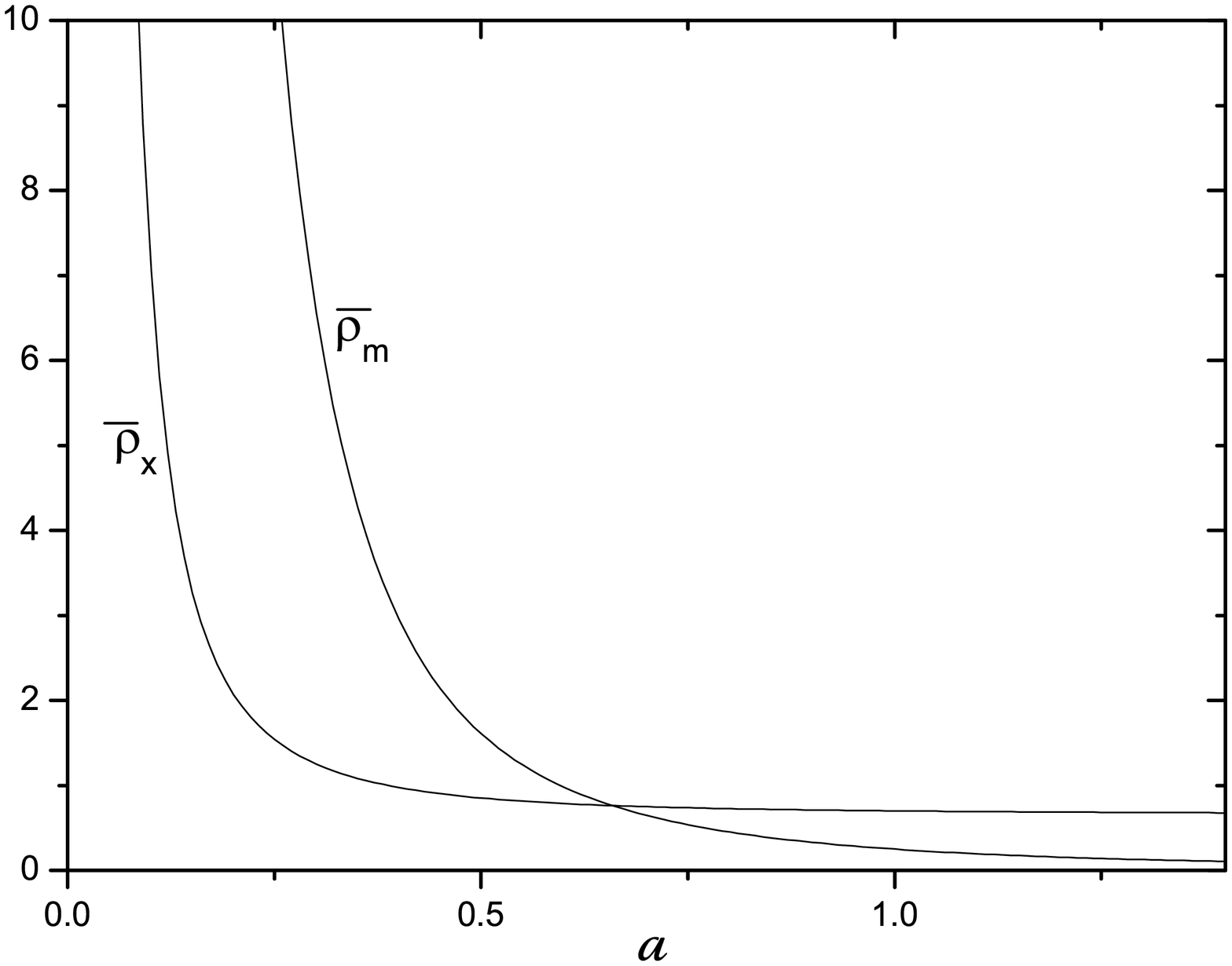}
  \end{minipage}
  \caption{{\small Left panel: Numerical solution to Eq. (\ref{eq:rhomevol4}) in terms of the
   dimensionless quantity $\bar{\rho}_{m} \equiv \kappa
    \rho_{m}/(3H^{2}_{0})$ with $\kappa \equiv 8 \pi G$, for the
     choice $\lambda=8$. Also shown is the curve of $\bar{\rho}_{x} = \bar{\rho}_{m}/r$.
     Right panel: The same but for the choice $\lambda=10$.}}
  \label{fig:rhomx}
\end{figure}

\subsection{ Third parametrization}
\noindent Similar  to the previous one is
\begin{equation}
r = r_{-} \, \exp \left(\frac{\lambda}{(1+a)^{2}}\right) \, ,
\label{eq:ra(a)5a}
\end{equation}
where $\lambda >0$ must be lower than $\simeq 10.125$ for $Q$ to
be positive and alleviate the coincidence problem. Here, $\, r_{-}
= r_{0}\, e^{-\lambda/4}$. Note that $\, r_{+} \, $ cannot be
arbitrarily large because it fulfills $r_{+} = r_{0} \, e^{3
\lambda/4}$.
\\  \

\begin{figure}[!htb]
  \begin{center}
    \begin{tabular}{c}
      \resizebox{120mm}{!}{\includegraphics{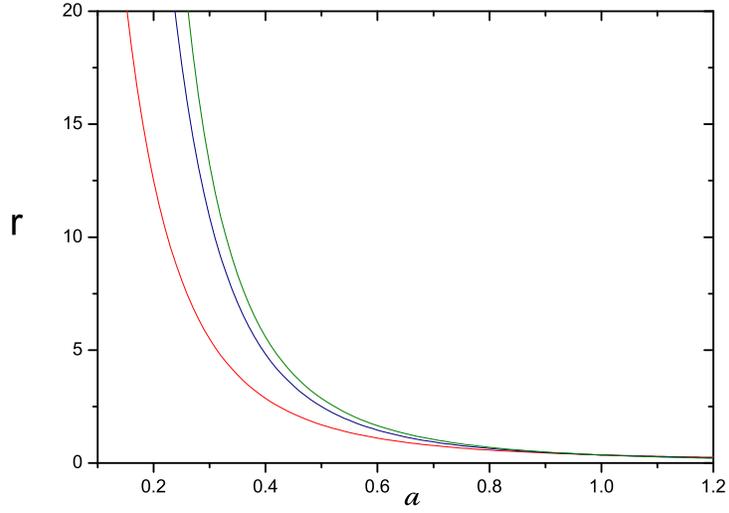}}\\
    \end{tabular}
    \caption{Plot of the ratio $\rho_{m}/\rho_{x}$ as a function of the scale factor as
    given by Eq. (\ref{eq:ra(a)5a}).
    The first two curves from left to right are for $\lambda = 8$, $\, \lambda =
    10$, respectively, with $\, r_{0} = 25/70$. Also shown for comparison is the
    analogous curve for the $\Lambda$CDM model with the same value for $r_{0}$.}
    \label{fig:ra(a)5a}
 \end{center}
\end{figure}

\begin{figure}[!htb]
  \begin{center}
    \begin{tabular}{c}
      \resizebox{120mm}{!}{\includegraphics{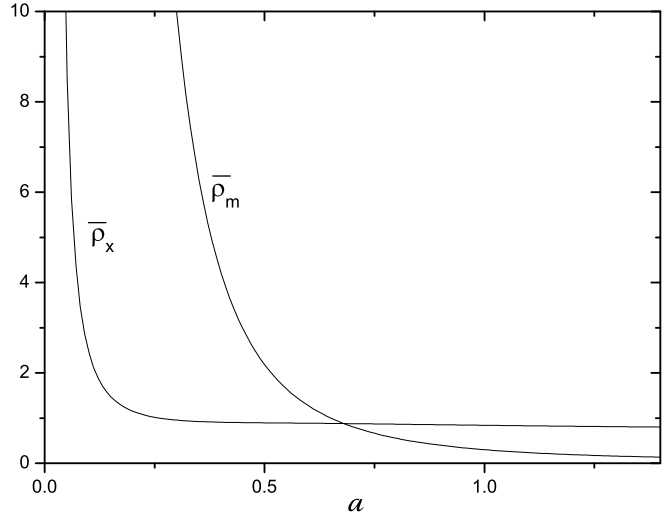}}\\
    \end{tabular}
    \caption{Numerical solution of Eq. (\ref{eq:rhomevol5}) in terms of the
    dimensionless quantity $\bar{\rho}_{m} \equiv \kappa
    \rho_{m}/(3H^{2}_{0})$ with $\kappa \equiv 8 \pi G$
    for the choice $\lambda = 9.89 $ and $\, r_{-} = 0.031334$.
    Also shown is the corresponding curve of $\bar{\rho}_{x}$.}
    \label{fig:numsp3}
 \end{center}
\end{figure}

\noindent Figure \ref{fig:ra(a)5a} depicts the ratio in terms of
the scale factor for $\lambda = 8$ and $\lambda =10$ as well as
the corresponding plot for the $\Lambda$CDM model.
\\  \
\noindent The interaction term in this case, with $w = -1$, reads
\begin{equation}
Q = \left(3 \, - \, \frac{2 \lambda a}{(1+a)^{3}}\right)\, H \,
\frac{\rho_{m}}{1\, +\, r_{-} \, e^{\lambda/(1+a)^2}} \, .
\label{eq:Q3rdparm}
\end{equation}
\\  \
\noindent Thus, the differential equation that governs  the
evolution of the DM energy density , assuming $w= -1$, reads
\begin{equation}
a\, \rho'_{m} \, +\, \left\{3\, + \, \left[\frac{2 \, \lambda \,
a}{(1+a)^{3}} \, - \, 3 \right]\, \frac{1}{1\, + \, r_{-} \, \,
e^{\lambda/(1+a)^{2}}} \right\} \, \rho_{m} = 0 \, .
\label{eq:rhomevol5}
\end{equation}
In the limit situations $a \rightarrow 0$ and $a \rightarrow
\infty$, one has $\rho_{m} \propto a^{-3r_{+}/(1+r_{+})}$ and $
\rho_{m} \propto a^{-3r_{-}/(1+r_{-})}$, respectively. A plot of
the numerical solution for $\lambda = 9.89$ is shown in Fig.
\ref{fig:numsp3}.

\section{Constraining the free parameters}
\noindent Thus far, we did not resort to any specific theory of
gravity. We have just used the FRW metric and the conservation
equations (1)-(3). Accordingly, the parameters entering the above
parametrizations have been left largely unconstrained. In order to
most effectively restrict them one should use some gravity theory
alongside the full set of observational data of supernovae type
Ia, baryon acoustic oscillations, history of the Hubble factor,
cosmic microwave background, growth factor, and so on. However,
this is beyond the scope of our work as we only aim to illustrate
how reasonable expressions for $Q$ can be obtained from the
qualitative evolution of the ratio $\rho_{m}/\rho_{x}$. This is
why we shall, for simplicity, assume $w = -1 $ throughout and
limit ourselves to quickly find acceptable values for the
remaining parameters. This will be done by using Einstein gravity
and the redshift at which the transition from decelerated to
accelerated expansion took place.
\\  \

\noindent Recently, Farooq and Ratra \cite{farooq2013}, based on
28 independent measurements of $H(z)$, determined  with
unprecedented accuracy  that redshift, namely, $z_{da} = 0.74 \pm
0.05$ ($1\sigma)$. The corresponding scale factor, $a_{da} =
(1+z_{da})^{-1}$, lies in the interval $0.56 \leq a_{da} \leq
0.59\,$, being $\, 0.57 \,$ the central value.
\\  \
From the deceleration parameter, $q = -\ddot{a}/aH^{2}$, and the
Einstein field equations for a spatially flat universe,
\begin{equation}\label{eq.efield}
H^{2} = \frac{\kappa}{3}\, \rho \, , \quad {\rm and} \quad
\frac{\ddot{a}}{a} = - \frac{\kappa}{3}\, (\rho\, + \, 3p),
\end{equation}
where $\rho = \rho_{b}\, + \, \rho_{m} + \rho_{x}$, $p = p_{x} = w
\rho_{x}$, we obtain
\begin{equation}\label{eq:q(r)1}
q = \frac{1}{2}\, \left[\frac{[(\rho_{b0}/\rho_{m0})+1]r\, + \, 1
\, + \, 3w}{[(\rho_{b0}/\rho_{m0})+1]r\, + \, 1} \right] \, .
\end{equation}
In the limiting cases $r \gg 1$ (i.e., $a \ll 1$, matter-dominated
epoch) and $r \ll 1 $ ($a \gg 1$, far future), one follows $q
=1/2$ and $q \rightarrow (1+3w)/2$, respectively, as it should. In
the particular case $w = -1$, it reduces to
\begin{equation}\label{eq:q(r)2}
q = \frac{1}{2}\,\left(\frac{1.2 \, r  \, - \, 2}{1.2 \, r\, +\,
1} \right) \, ,
\end{equation}
where we have considered $\rho_{b0}/\rho_{m0} = 0.2$.
\\  \

\noindent The latter expression of $q$ together with the above
value of $a_{da}$ can be employed to get acceptable values for the
free parameters in Eqs. (\ref {eq:r(a)1}), (\ref{eq:ra(a)5}), and
(\ref{eq:ra(a)5a}).
\\   \

\begin{figure}[!htb]
  \begin{center}
    \begin{tabular}{c}
      \resizebox{120mm}{!}{\includegraphics{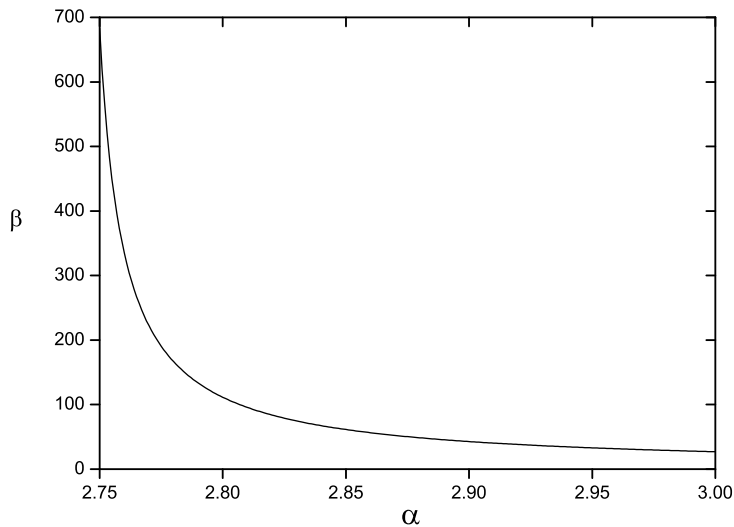}}\\
    \end{tabular}
    \caption{The parameter $\beta$ in terms of $\alpha$ as imposed by the constraint
    $q(a=0.57) = 0$ on the first parametrization, Eq. (\ref{eq:r(a)1}), with $r_{-} = 0$.}
    \label{fig1pbalpha}
 \end{center}
 \end{figure}

\noindent For the first parametrization Eq.  (\ref {eq:r(a)1}),
Fig. \ref{fig1pbalpha} depicts the dependence imposed on the
parameters $\beta \, $ and $\, \alpha$ by the observational
constraint $q(a_{da}) = 0$. For simplicity we have taken $r_{-} =
0$.

\noindent Then, the deceleration parameter vanishes at $a_{da} =
0.57$ for  any point in the graph. Since we are dealing with the
fixing of two parameters a further fixed point is necessary. We
take the product $t_{0}\, H_{0}$ where $t_{0}$ and $H_{0}$ stand
for the age of the Universe and the Hubble constant, respectively.
Measurements of the age of globular clusters and other old objects
place the former in the interval $12 \lesssim t_{0} \lesssim 14\,
$ Gyr \cite{ages}. As for the latter, it is safe to say after the
results of the Planck Collaboration \cite{Ade2013}, $H_{0} = 67.3
\pm 1.2 \,$ Km/s/Mpc,  and the local measurements of Riess {\it et
al} \cite{Riess2011}, $H_{0} = 73.8 \pm 2.4\, $ Km/s/Mpc, that the
said product must lie in the range $0.81 \lesssim t_{0} \, H_{0}
\lesssim 1.09$. Here,
\begin{equation}
t_{0} \, H_{0} = \int_{0}^{1}{\frac{{\rm d}a}{a \,
\sqrt{\Omega_{b0}\, a^{-3} \, + \, \Omega_{m0} \, f(a)\, + \, (1-
\Omega_{b0} - \Omega_{m0}) \, g(a)}}} \, , \label{eq:t0H0}
\end{equation}
where $f(a) \equiv \rho_{m}/\rho_{0}$ is given by the right-hand
side of Eq. (\ref{eq:rhom1}) with $r_{-}=0$ and $r_{+} =
r_{0}(1+\beta)$, $g(a) \equiv \rho_{x}/\rho_{x0} = r_{0} \,
f(a)/r$, $\Omega_{b0} = 0.05$, and $\Omega_{m0} = 0.25$. Thus,
only that pair of values $(\alpha, \beta)$ lying on the graph of
Fig. \ref{fig1pbalpha} such that the corresponding product $t_{0}
\, H_{0}$ lies in the above interval can be considered for the
parametrization. There is a further qualitative restriction. In
order that dark energy does not spoil the standard picture of
primeval nucleosynthesis it must be present only in a
comparatively small amount at that epoch, namely, $\Omega_{x} <
0.045$ \cite{prd_bean}. This implies that $\beta$ must be large,
no lower than, say, $50$. At any rate, given the absence of a
second strict observational constraint at the background level,
there is a certain latitude in choosing the values of both
parameters, i.e., properly speaking there is not a ``best fit" as
a handful pair of values ($\alpha, \beta$) fits the data equally
well. Thus, only after a substantial improvement on the
observational estimation of $H_{0}$ and $t_{0}$, it will be
possible to determine the best fit. At any rate, for the sake of
illustrating the method, we take $\alpha = 2.8$ and $\beta =
111.33$ (consequently, $r_{+} \simeq 40.12$), and this corresponds
to $\, t_{0} \, H_{0} \simeq 0.98$.
\\  \

\noindent Likewise, for the second parametrization setting
$q(a_{da}) = 0$ leads to $\lambda = 11.25$, thereby $r_{-} \simeq
0.0013$ and $r_{+} \simeq 112$. Similarly, for the third
parametrization one obtains $\lambda = 9.89$ whence $r_{-} \simeq
0.030$ and $r_{+} \simeq 594$. As it can be numerically checked,
the product $t_{0}\, H_{0}$ yields $0.99$ and $0.98$ for the
second and third parametrizations, respectively. Thus, in both
cases it falls within the allowed range.

\begin{figure}[!htb]
  \begin{center}
    \begin{tabular}{c}
      \resizebox{120mm}{!}{\includegraphics{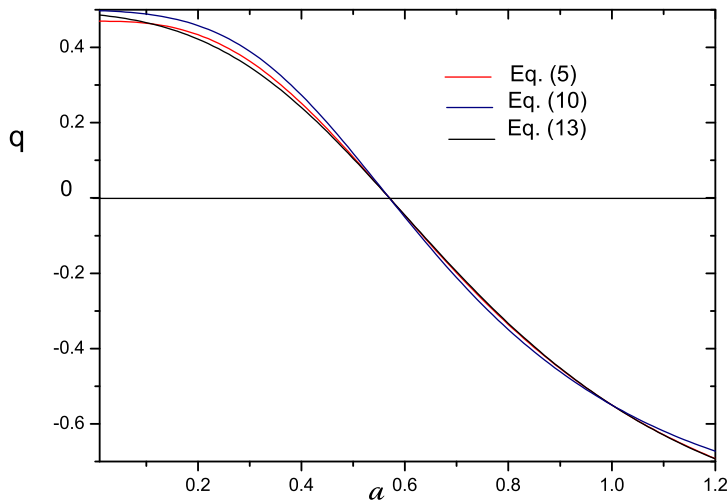}}\\
    \end{tabular}
    \caption{{\small Evolution of the deceleration parameter in terms of the scale factor
    for the first, second and third parametrizations.  In drawing the curves, we
    used $\, w = -1$, $r_{0} = 25/70 $, and $\, \rho_{b0}/\rho_{m0}
    =0.2$.}}
    \label{fig:q(a)123}
 \end{center}
\end{figure}

\noindent Figure \ref{fig:q(a)123} collects the evolution of the
deceleration parameter for the three parametrizations [Eqs.
(\ref{eq:r(a)1}), (\ref{eq:ra(a)5}), and (\ref{eq:ra(a)5a}),
respectively] for the above parameter values. Though not shown in
the figure, $q(r_{+}) \simeq 0.46$ for the first one and $\simeq
0.49$ for the other two. Likewise, $q(r_{-}) \simeq -1$ for the
two first parametrizations and $ \simeq -0.95$ for the third one.

\noindent To establish whether the above parametrizations,
alongside the parameter values just obtained by imposing
$q(a_{da}) = 0$, are acceptable \textemdash at least at a good
qualitative level \textemdash $\;$ we next determine $H(z)$ for
each parametrization and contrast it graphically with the history
of the Hubble factor.
\\  \
\noindent From the expressions $q = \ddot{a}/aH^{2}$ and $H =
\dot{a}/a$, it follows
\begin{equation}\label{eq:H(z)}
H= H_{0} \, \exp\left\{\int_{0}^{z}{[1 \, + \, q(x)]\, \, {\rm d}
\ln(1+x)} \right\} \, ,
\end{equation}
where $H_{0}$ is Hubble's constant. Expression (\ref{eq:H(z)})
provides us with the Hubble history for every parametrization of
$\, \rho_{m}/\rho_{x} \,$ via the corresponding deceleration
parameter.

\begin{figure}[!htb]
  \begin{center}
    \begin{tabular}{c}
      \resizebox{120mm}{!}{\includegraphics{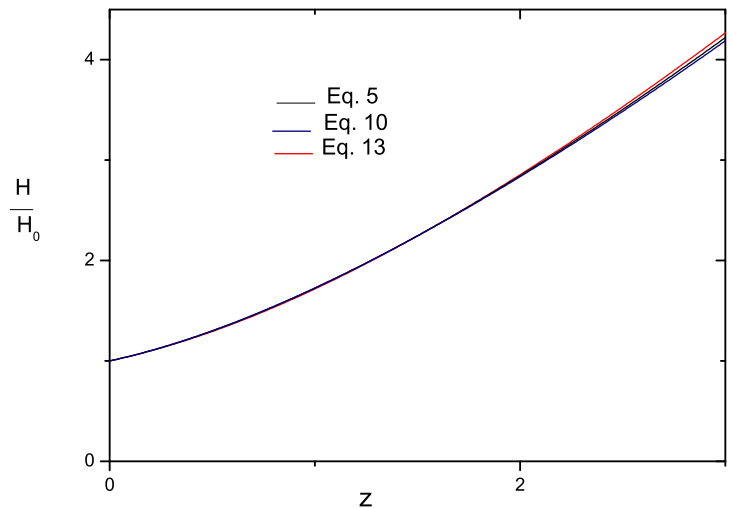}}\\
    \end{tabular}
    \caption{{\small Hubble history corresponding to the three
    parametrizations, Eqs. (\ref{eq:r(a)1}), (\ref{eq:ra(a)5}), and
    (\ref{eq:ra(a)5a}). For the curve corresponding to the first parametrization
    Eq. (\ref{eq:r(a)1}), we have assumed $\alpha = 2.5$ and $\beta = 697/3$, for the
    second parametrization Eq. (\ref{eq:ra(a)5}), $\lambda = 8$, and for the
    third prametrization Eq. (\ref{eq:ra(a)5a}) $\lambda = 9.89$.
    In drawing the curves we have taken $\, w = -1$, $\, r_{0} =25/70$,
    and $\, \rho_{b0}/\rho_{m0} =0.2$.}}
    \label{fig:H(z)123}
 \end{center}
\end{figure}

\begin{figure}[!htb]
  \begin{center}
    \begin{tabular}{c}
      \resizebox{120mm}{!}{\includegraphics{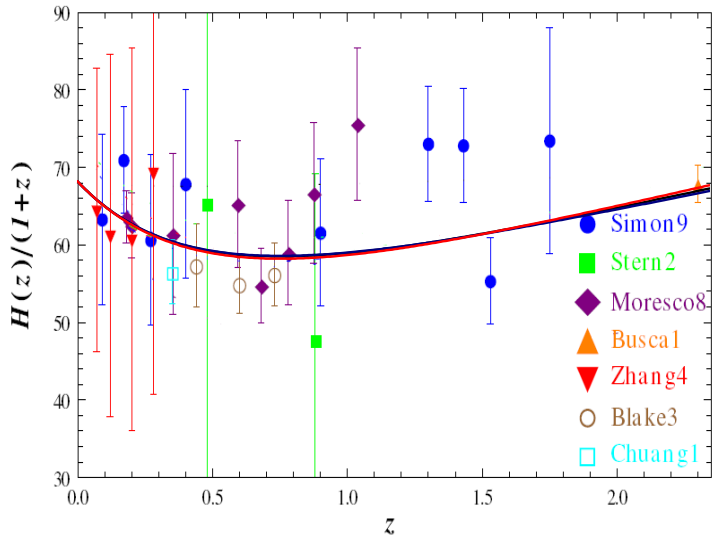}}\\
    \end{tabular}
    \vspace{-6.0cm \caption{{\small Plot of the ratio $H(z)/(1+z)$ assuming
    $H_{0} = 68\, {\rm Km/(s \cdot Mpc)}$. Black, blue, and red lines correspond
    to the first, second and third parametrizations [Eqs. (\ref{eq:r(a)1}), (\ref{eq:ra(a)5}),
    and (\ref{eq:ra(a)5a})], respectively. The data error bars indicate $1 \sigma$ confidence level.
    The $H(z)$ data were borrowed from Refs. \cite{busca2012} - \cite{zhang2012}.}}
    \label{fig:farooq1}}
 \end{center}
\end{figure}

\begin{figure}[!htb]
  \begin{center}
    \begin{tabular}{c}
      \resizebox{120mm}{!}{\includegraphics{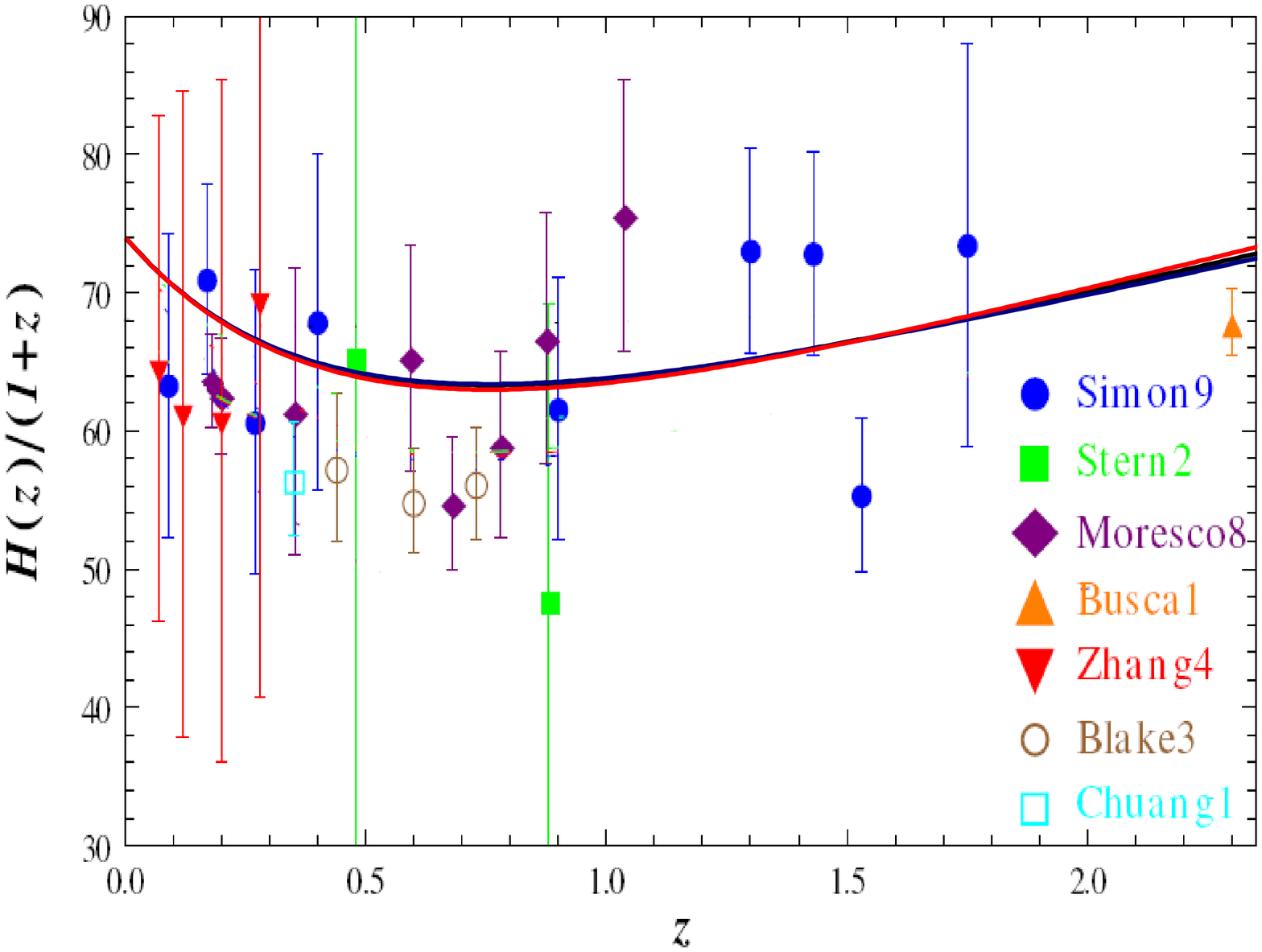}}\\
    \end{tabular}
   \vspace{-5.0cm \caption{{\small The same as Fig. \ref{fig:farooq1} but
using $H_{0} = 73.8 \, {\rm Km/(s \cdot
    Mpc)}$.}}
    \label{fig:farooq2}}
 \end{center}
\end{figure}

\noindent Figure \ref{fig:H(z)123} depicts the Hubble history
(normalized to $H_{0}$) in the redshift  interval $0 < z < 3$ for
the three parametrizations: (\ref{eq:r(a)1}), (\ref{eq:ra(a)5}),
and (\ref{eq:ra(a)5a}). They begin to differ only after $z = 2$.
Likewise, Figs. \ref{fig:farooq1} and \ref{fig:farooq2} show the
ratio $H(z)/(1+z)$ for $0 < z < 24$, for the same three
parametrizations confronted with the 28 observational $H(z)$ data
collected in \cite{farooq2013}. In Fig. \ref{fig:farooq1} we have
used $H_{0} = 68 \, {\rm Km/(s \cdot Mpc)}$ \cite{chen2003}, a
value within $1\sigma$ of the one reported by Ade {\it et al.},
namely, $67.3 \pm 1.2$ in the same units \cite{Ade2013}. In Fig.
\ref{fig:farooq2}, we  used the central value $H_{0}= 73.8 \, {\rm
Km/(s \cdot Mpc)}$, obtained by Riess {\it et al.} by local
measurements \cite{Riess2011}. The graph for the $\Lambda$CDM
model is not shown because it essentially overlaps the other
three.
\subsection{The relative weight of $Q$}
\noindent As mentioned in Sec. II, we do not expect  the
interaction term $Q$ to be larger than the second term in the
conservation equation for the DM energy density, Eq. (2). Then,
assuming $w = -1$, every parametrization should fulfill
\begin{equation}\label{eq:zetaratio}
\zeta \equiv \frac{3\, \rho_{m} \, H}{Q} = \frac{3 \, (1+r)}{3\, +
\, (a \, r'/r)} > 1.
\end{equation}
\\  \

\begin{figure}[!htb]
  \begin{center}
    \begin{tabular}{c}
      \resizebox{120mm}{!}{\includegraphics{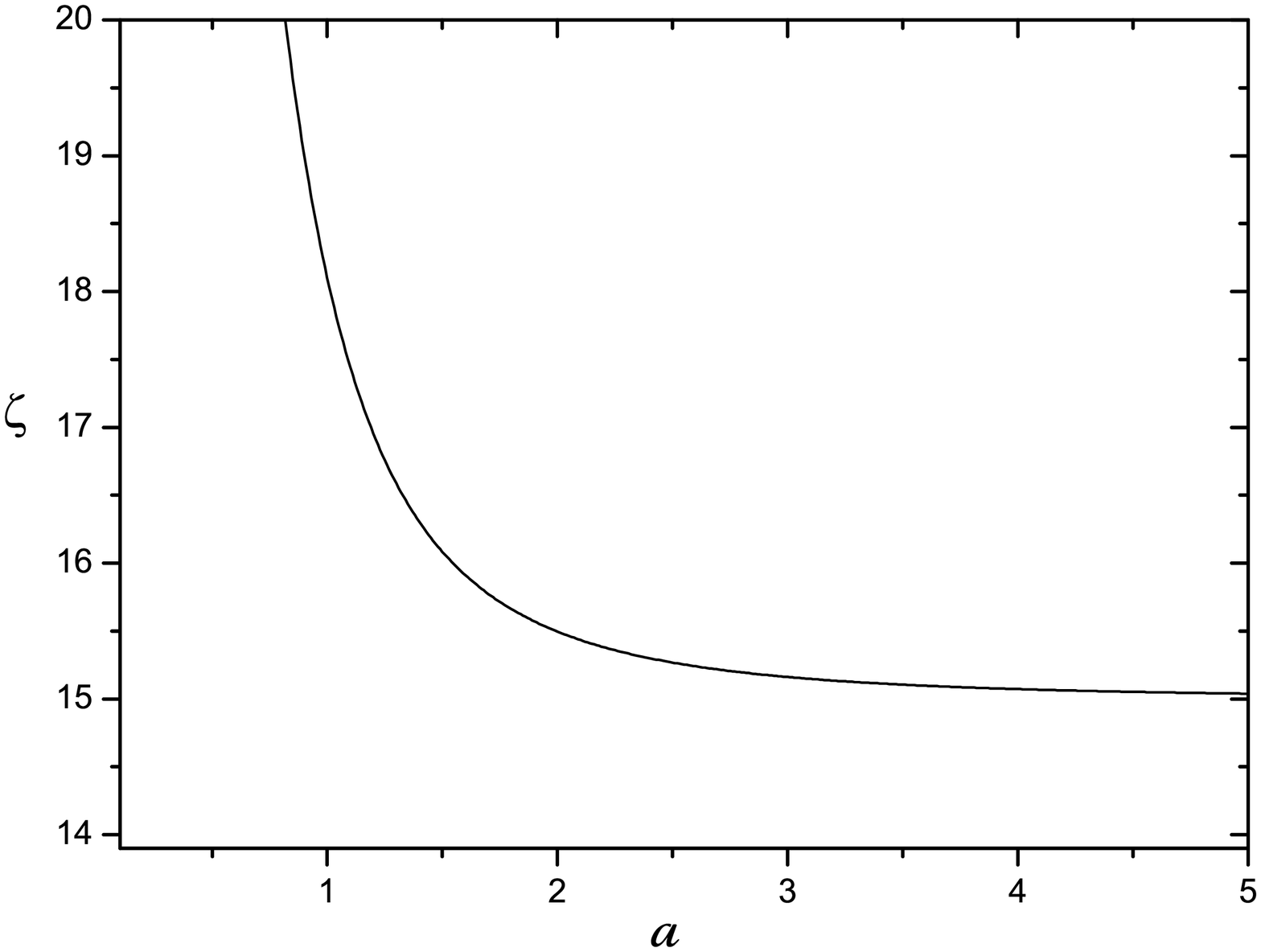}}\\
    \end{tabular}
    \caption{{\small Evolution of the ratio (\ref{eq:zetaratio}) with expansion for the first parametrization,
   Eq. (\ref{eq:r(a)1}). The larger the scale factor, the smaller $\zeta$. In drawing the curve we have
   used $\, r_{0} = 25/70$, $\, \beta = 111.33 $, and $\, \alpha = 2.8$.}}
    \label{fig:zetap1}
 \end{center}
\end{figure}

\noindent This is satisfied by the three parametrizations
\textemdash Eqs. (\ref{eq:r(a)1}), (\ref{eq:ra(a)5}) and
(\ref{eq:ra(a)5a}). For the first one, Fig. \ref{fig:zetap1}
illustrates (using the parameter values derived in the previous
section) how the relative importance of $Q$, i.e., ($\zeta^{-1}$),
increases with expansion and is bounded from above to be lower
than unity (in fact, in this case, $\zeta \geq 15$ for the whole
cosmic history).
Notice that the graph is fully consistent with the fact that $\mid
\rho'_{m} \mid $ is much larger for $a \ll 1 $ than for $a$ around
and beyond unity (see Fig. \ref{fig:sol1}).
\\  \

\noindent A rather similar situation occurs for the second and
third parametrizations [Eqs. (\ref{eq:ra(a)5}) and
(\ref{eq:ra(a)5a})]. This is why we omit the corresponding graphs
of $\zeta$.

\section{Two ``possible" degeneracies}
\noindent In principle, any given ratio $r(a)$ may be equally
associated to two different scenarios; viz., the one considered so
far (a nonvanishing interaction term $\, Q \, $ with constant EOS
$\, w$ for the DE component) and also to the case of vanishing $\,
Q$ (no interaction) but with $w$ given by a suitable EOS,
different from the previous one. Therefore, one could not tell
whether such a ratio points to the existence of an interaction or,
perhaps, to a more involved EOS. In other words, a possible
degeneracy between $Q$ and $w$ arises. Here, we shall argue that
this degeneracy is only apparent because it can be readily broken
at the background level since observable quantities, such as the
Hubble factor and the deceleration parameter, significantly differ
from one scenario to the other.
\\ \

\noindent Indeed, by setting $Q = 0$ in the set of equations
(1)\textemdash(\ref{eq:conservation}), the EOS that produces the
same $r(a)$ as the one associated to a nonvanishing $Q$ is easily
found,
\begin{equation}
w(a) = - \frac{r'}{r} \frac{\rho'_{m}}{\rho_{m}}.
\label{eq:eos1}
\end{equation}
For illustrative purposes, let us focus on the second
parametrization of $r$, namely, (\ref{eq:ra(a)5}), and the
corresponding interaction term is given by (\ref{eq:Q2ndparm}).
With the help of (\ref{eq:rhomevol4}) we obtain
\begin{equation}
w(a) = - \frac{\lambda}{(1+a)^{2}} \, \, \frac{a}{3\, + \,
\left[\frac{\lambda \, a}{(1+a)^{2}} \, - \, 3 \right]\,
\frac{1}{1\, + \, r_{-} \, \, e^{\lambda/(1+a)}}} \, .
\label{eq:eos2}
\end{equation}

\begin{figure}[!htb]
  \begin{center}
    \begin{tabular}{c}
      \resizebox{120mm}{!}{\includegraphics{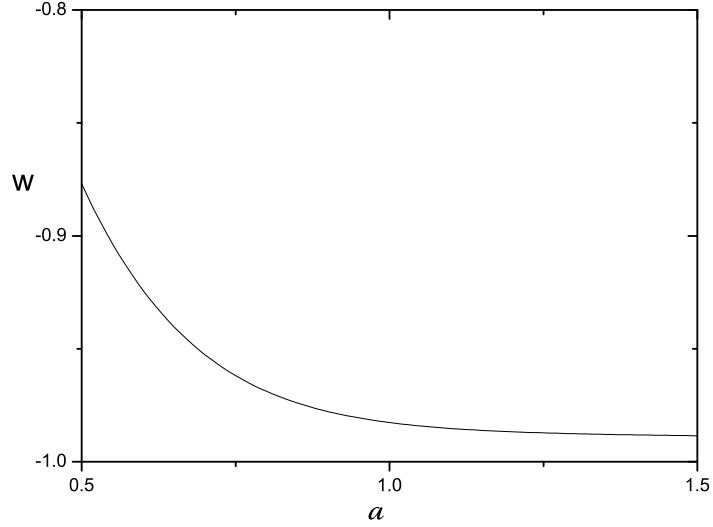}}\\
    \end{tabular}
    \caption{Evolution of the EOS corresponding to the second parametrization,
    Eq. (\ref{eq:ra(a)5}), assuming $Q = 0$. In drawing the curve we have taken $\lambda =
    11.25$ and $r_{-} = 0.0013$ (the best fit-values found above; see text).}
    \label{fig:wa}
 \end{center}
\end{figure}

\noindent Figure \ref{fig:wa} depicts the dependence of $w$ on the
scale factor. This variable EOS produces the same $r(a)$ as the
case with interaction, Eq. (\ref{eq:Q2ndparm}). However, this EOS
seems to be at variance with observation because it contrasts with
the findings of Serra {\it et al.} \cite{prd_serra}, who using a
variety of observational data conclude that the EOS, varies, if at
all, very little from $z = 1$ (i.e., $a = 0.5$) up to the present.
\\  \

\noindent Let us consider the Hubble factor for the scenario with
$Q =0$ and the EOS (\ref{eq:eos2}),
\begin{equation}
\frac{H(a)}{H_{0}} = \sqrt{\Omega_{b0} a^{-3}\, + \, \Omega_{m0}
a^{-3}\, + \, \Omega_{x0}\, \exp\left[- 3 \,
\int_{1}^{a}{\frac{1+w(x)}{x}\, {\rm d} x} \right]} \, ,
\label{eq:H(a)2}
\end{equation}
where  $\Omega_{b0} = 0.05$, $\Omega_{m0} = 0.25$, and
$\Omega_{x0} =0.70$ are the current fractional densities of the
various energy components.
\\  \

\begin{figure}[!htb]
  \begin{center}
    \begin{tabular}{c}
      \resizebox{120mm}{!}{\includegraphics{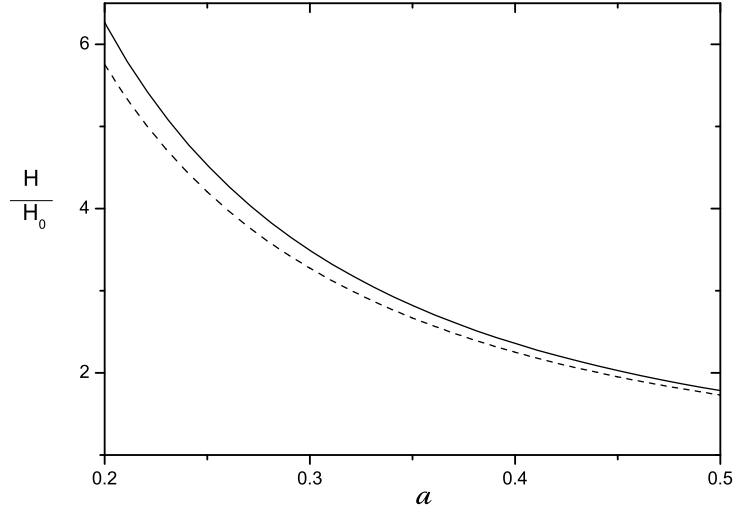}}\\
    \end{tabular}
    \caption{Evolution of the Hubble factor normalized to its present value, $H_{0}$. The solid line
    corresponds to Eq. (\ref{eq:H(a)2}) with $w(a)$ given by (\ref{eq:eos2}). The dashed line
    depicts the Hubble factor associated to the case of the same $r(a)$ but with $Q$ given
    by (\ref{eq:Q2ndparm}) and $w = -1$. In drawing both graphs we assumed $\Omega_{b0} = 0.05$,
    $\Omega_{m0} = 0.25$, $\Omega_{x0} = 1 - \Omega_{b0}- \Omega_{m0}$, $\lambda = 11.25$,
    and $r_{-} = 0.0013$.}
    \label{fig:H(a)comp2}
 \end{center}
\end{figure}

\noindent Figure \ref{fig:H(a)comp2} depicts the evolution of the
Hubble factor associated  to Eq. (\ref{eq:H(a)2}) (solid line),
and the one corresponding to interacting scenario with $w = -1$
(dashed line), both are normalized to $H_{0}$. Both graphs depart
more and more from each other as the scale factor decreases.
Further, since $H(a)$  significantly differs between both models,
the luminosity distance is also affected, as well as all other
cosmological observable quantities (e.g., supernovae Ia
brightness, age of the Universe, background acoustic oscillations,
etc.). So, the degeneracy gets already broken at the background
level.
\\  \

\begin{figure}[!htb]
  \begin{center}
    \begin{tabular}{c}
      \resizebox{120mm}{!}{\includegraphics{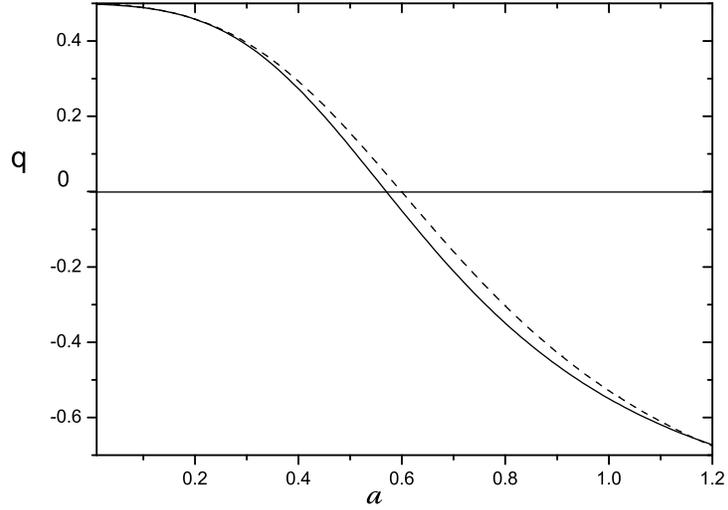}}\\
    \end{tabular}
    \caption{Evolution of the deceleration parameter for the model with $Q = 0$ (solid line)
    and for the model with $Q$ given by (\ref{eq:Q2ndparm}) (dashed line). Both models
    share the same ratio $r$, equation (\ref{eq:ra(a)5}), but their respective EOS are
    different -see text. In drawing the graphs we took  $\rho_{b0}/\rho_{m0} = 0.2$,
    $\lambda = 11.25$, and $r_{-} = 0.0013$.}
    \label{fig:q(a)resp}
 \end{center}
\end{figure}

\noindent The consideration of the deceleration parameter
strengthens this point further. Equation (\ref{eq:q(r)1})  applies
to both instances, viz., the model with $Q = 0$, and with $r$ and
$w$ given by Eqs. (\ref{eq:ra(a)5}) and (\ref{eq:eos2}),
respectively, and the model with $Q$ given by Eq.
(\ref{eq:Q2ndparm}), where $w$ is fixed to  $-1$. The observable
$q$ significantly varies from one model to the other because
$w(a)$, which enters (\ref{eq:q(r)1}), does also, while all the
other quantities in (\ref{eq:q(r)1}) are identical. Figure
\ref{fig:q(a)resp} makes this explicit.
The difference is already significat, though not large, for $a =
0.57$ (the value observationally determined by Farooq and Ratra
\cite{farooq2013} at which $q$ vanishes). For the $Q = 0$ model,
$q(a = 0.57) \simeq 0.046$, while for the other model $q(a = 0.57)
\simeq 0.0031$, with the former figure being about $15$ times the
latter. Again, the degeneracy is readily broken at the background
level.
\\  \

\noindent Here, we have just considered for illustrative purposes
the interaction term associated to the second parametrization. An
identical conclusion can be attained for the other interaction
terms above.
\\  \
\bigskip

\noindent Another apparent degeneracy is as follows. In principle,
one could think that it would be possible to reproduce the same
expansion history $H(z)$ and, consequently, the same deceleration
parameter $q(z)$ in a noninteracting scenario ($Q = 0$) by using a
suitably tailored EOS $w(z)$, for the dark energy component  in an
interacting scenario ($Q \neq 0$) with $w =$ constant. Obviously,
now the ratio $r(z)$ will differ from one scenario  to the other.
Since $H(z)$ is the same in both, they will share the same
luminosity and angular distances whence the models will be
indistinguishable at the background level.
\\   \

\noindent However, as we will see, this possibility does not exist
in general, nor does it exist in the particular cases of the three
parametrizations above or for parametrizations (\ref{eq:r(a)Q1})
and (\ref{eq:r(a)Q2}) considered below. Indeed, by equating the
right-hand sides of Friedmann's equation for spatially flat
universes corresponding to the interacting model, and assuming $w
= -1$ (for simplicity), with the nointeracting one $Q = 0$ but
with an undetermined EOS $w(a)$, it follows that the latter
quantity is given implicitly by the expression
\begin{equation}
\int_{a}^{1}{\frac{1 \, + \, w(a)}{a}\, {\rm d}a} =
\textstyle{1\over{3}}\, \ln \left\{r_{0} \, \left[f(a) \, - \,
a^{-3}\right] \, + \, g(a)\right\}, \label{Eq:w(a)}
\end{equation}
where $f(a)$ stands for the DM ratio $\rho_{m}(a)/\rho_{m0}$ in
the interacting model [e.g. by the right-hand side of Eq.
(\ref{eq:rhom1}) in the case of the first parametrization] and
$g(a) \equiv \rho_{x}/\rho_{x0} = r_{0} f(a)/r(a)$. In writing Eq.
(\ref{Eq:w(a)}), we bore in mind that the current fractional
densities of the three components coincide in both models.
\\   \

\begin{figure}[!htb]
  \begin{center}
    \begin{tabular}{c}
      \resizebox{120mm}{!}{\includegraphics{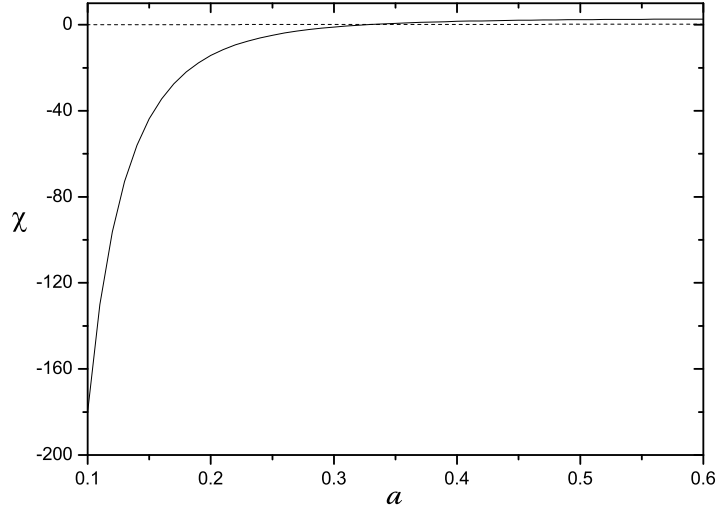}}\\
    \end{tabular}
    \caption{Evolution of function $\chi$,  Eq. (\ref{eq:chi(a)}), in terms of the scale
    factor for the case of the first parametrization, Eq. (\ref{eq:r(a)1}).
    As is apparent, $\chi > 0$ only for $ \, a > 0.4$, thereby no EOS of any
    noninteracting model can reproduce the Hubble history, $H(z)$, of the model corresponding to
    the first parametrization. In plotting the graph we have used $\Omega_{b0} = 0.05$,
    $\Omega_{m0} = 0.25$, $r_{-} = 0$, $r_{+} = 40.12$, $\alpha = 2.8$, and $\beta =111.33$.}
    \label{fig:chi(a)p1}
 \end{center}
\end{figure}

\begin{figure}[!htb]
  \begin{center}
    \begin{tabular}{c}
      \resizebox{120mm}{!}{\includegraphics{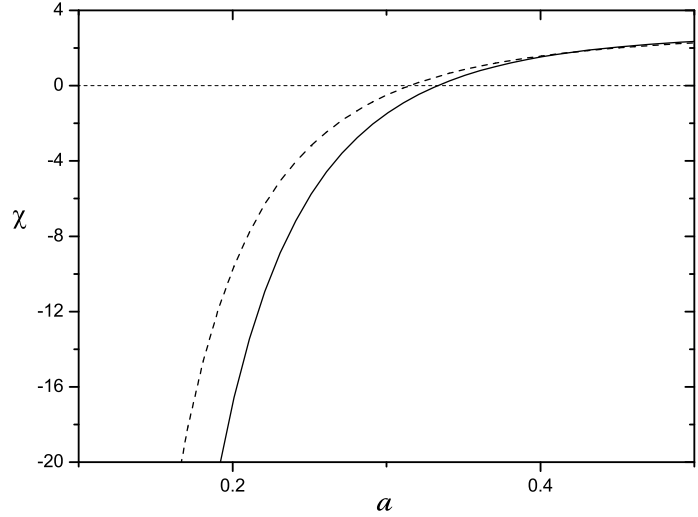}}\\
    \end{tabular}
    \caption{Solid line: Evolution of function $\chi$, Eq. (\ref{eq:chi(a)}), in terms of the scale
    factor for the case of the second parametrization, Eq. (\ref{eq:ra(a)5}). Dashed line: The
    same but for the second parametrization, Eq.
    (\ref{eq:ra(a)5a}). As is seen, $\chi(a \lesssim 0.32) <0$ and $\chi(a \lesssim 0.30)<0$
    for the second and third parametrizations, respectively.
    In drawing the graphs we have used the value of the parameter $\lambda$
    that best fits the observational data [$q(a= 0.57) = 0$], i.e., $11.25$ and
    $9.89$ for the second and third parametrizations, respectively.
    In both cases we have assumed $\Omega_{b0} = 0.05$ and $\Omega_{m0} = 0.25$.}
    \label{fig:chi(a)p23}
 \end{center}
\end{figure}

\noindent Because of at any $a < 1$, the amount of DM in the
interacting model is necessarily less than in the case of
vanishing $Q$ (in which instance the DM density would obey
$\rho_{m} = \rho_{m0}\, a^{-3}$), one follows $f(a) < a^{-3}$, and
the curly brackets on the right hand-side of (\ref{Eq:w(a)}) can
be negative in some interval of the scale factor when $a < 1$.
This severely constrains $w(a)$ since the said brackets  must be
non-negative for any value of $\, a$ if $w(a)$ is to exist at all.
Accordingly, if the function
\begin{equation}
\chi(a) \equiv f(a)\, - \, a^{-3}\, + \, \frac{f(a)}{r(a)}
\label{eq:chi(a)}
\end{equation}
takes negative values in some interval of the scale factor, there
will be no  $w(a)$ capable of reproducing the $H(a)$ function of a
given interacting model. This is the case of the three
parametrizations (\ref{eq:r(a)1}), (\ref{eq:ra(a)5}), and
(\ref{eq:ra(a)5a}), as Figs. \ref{fig:chi(a)p1} and
\ref{fig:chi(a)p23} illustrate.
\\  \

\begin{figure}[!htb]
  \begin{center}
    \begin{tabular}{c}
      \resizebox{120mm}{!}{\includegraphics{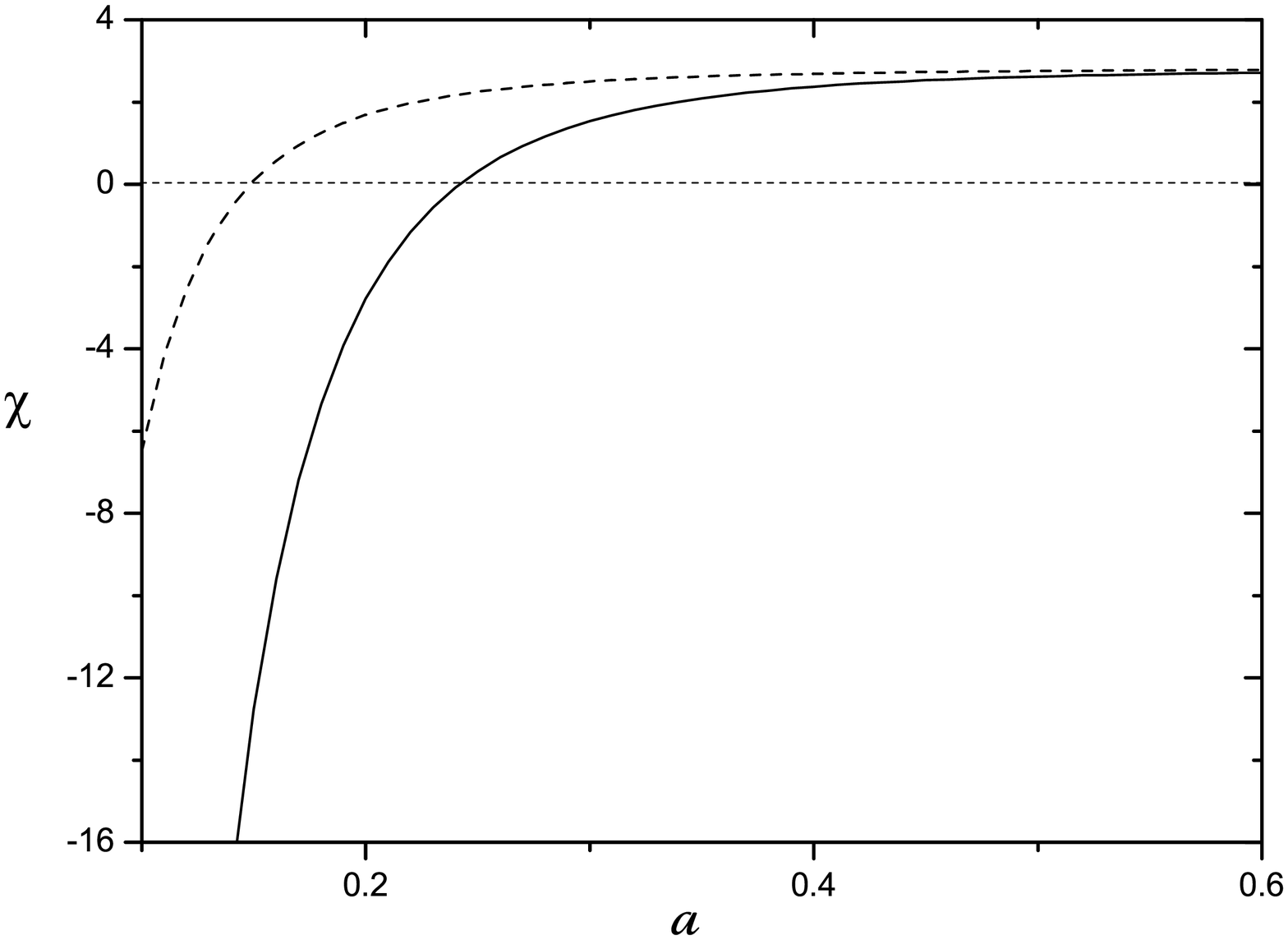}}\\
    \end{tabular}
    \caption{Solid line: Evolution of function $\chi$, Eq. (\ref{eq:chi(a)}), in terms of the scale
    factor for the case of  parametrization (\ref{eq:r(a)Q1}). Dashed line: The
    same but for  parametrization (\ref{eq:r(a)Q2}). For both
    parametrizations  $\chi$ is negative in a considerable interval of the scale factor. In
    drawing the graphs we have considered $r_{0} = 0.25/0.70$, $w = -1$, and $\xi = 0.01$.
    In both cases we have assumed $\Omega_{b0} = 0.05$ and $\Omega_{m0} = 0.25$.}
    \label{fig:chi(a)Q1Q2}
 \end{center}
\end{figure}

\noindent As Fig. \ref{fig:chi(a)Q1Q2} shows, the same is true for
the parametrizations given by Eqs. (\ref{eq:r(a)Q1}) and
(\ref{eq:r(a)Q2}) (see below). These  correspond to the widely
used interaction terms $Q = \xi H \rho_{m}$ and $Q = \xi H
\rho_{x}$, respectively. Here, $\xi$ is a semipositive, small,
constant parameter.
\\  \

\noindent In general, it is reasonable to expect that for any $\,
Q >0 \,$ interaction the  $\, f(a)\,$ function may be ``piecewise"
expressed as $\, f(a) = a^{-3+\epsilon} \,$ with  $\, 0 < \epsilon
\ll 1 \, $ a constant  parameter in every small-scale factor
interval but such that it slightly varies between two neighboring
intervals; the smaller the interval, the better the approximation
is. Then, the expression
\begin{equation}
\chi (a) \simeq a^{-3} \, \left[a^{\epsilon} \left(1+
\frac{1}{r(a)} \right)\, - \, 1\right]
 \label{eq:chi3}
\end{equation}
will be valid in every small interval of the scale factor.
Accordingly, bearing in mind that ${\rm d}r(a)/{\rm d}a < 0 $, and
that  for $r \gtrsim (0.25/0.70)^{1/3} \simeq 0.71$, DM dominates
over DE (otherwise, the large-scale structure we observe today
could not be accounted for), it follows that $\, r(a) > 1 \,$ for
$\, a \,$ small enough. Then, for some scale factor, say $a_{*} <
1$, $\, \chi(a)$ will be negative in the interval $0 < a < a_{*}$,
and no EOS $w(a)$ will exist such that the corresponding
noninteracting model can reproduce the same history of the Hubble
function of a given interacting model with constant $w$.
\\  \

\bigskip

\begin{figure}[!htb]
  \begin{center}
    \begin{tabular}{c}
      \resizebox{120mm}{!}{\includegraphics{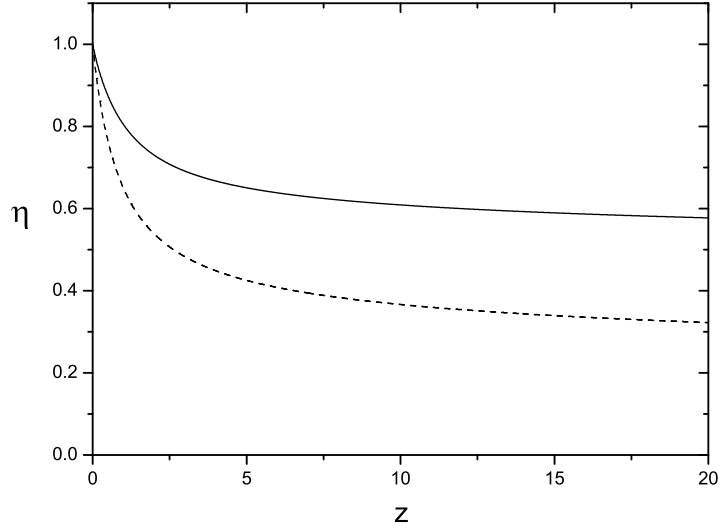}}\\
    \end{tabular}
    \caption{Evolution of the ratio $\eta \equiv
\rho_{m}(z)/\rho_{m0}\, (1+z)^{3}$, where $\rho_{m}(z)$ is the
dark matter density in the case of the second parametrization. The
solid line corresponds to $\lambda = 10$, the dashed one to
$\lambda = 8$. In drawing both curves  $r_{0} = 25/70$ was
assumed.}
    \label{fig:rhomz}
 \end{center}
\end{figure}

\noindent On the other hand, even wildly assuming the existence of
such EOS, both kinds of models will be readily told apart at the
perturbative level without the need to perform a detailed
analysis. Indeed, because in the interacting model DM is
continuously added to that component from the DE (recall that $Q >
0$), as mentioned above,  the amount of dark matter in the past
was necessarily less than in noninteracting scenarios at the same
redshift. This is illustrated by Fig. \ref{fig:rhomz} where the
ratio $\eta \equiv \rho_{m}(z)/\rho_{m0}\, (1+z)^{3}$ vs. $z$ is
depicted for the case of the second parametrization, Eq.
(\ref{eq:ra(a)5}), in the redshift interval $0 \leq z \leq 20$. On
the other hand, since the present amount of matter (baryonic,
$\Omega_{b0}$ and dark $\Omega_{m0}$) has to be the same in both
models, the growth function, $ f_{g} = {\rm d} \ln D_{+}/ {\rm d}
\ln a$ in the interacting scenario must be larger at any redshift
than in the $\Lambda$CDM model, while the same function in the
noninteracting scenario will be, at most, as large as in the
latter model. While current measurements of $f_{g}(z)$ are not
accurate enough to do the job, future measurements \textemdash
possibly the next generation of data\textemdash  will, hopefully,
be precise enough to discriminate between the said scenarios. This
may  well be achieved by the Euclid mission \cite{euclid}. Thanks
to the latter, $f_{g}(z) \,$ is expected to be determined   with
an accuracy between $\, 1\% \, $ and $\, 2.5\% $  in the redshift
interval $\, 0.5 < z < 2$ (see Figs. 2 and 3 in Di Porto {\it et
al.} \cite{DiPorto} and Figs. 14 and 15 in Amendola {\it et al.}
\cite{Luca}). In summary, if, by any chance, an interacting model
and noninteracting one happen to share an identical expansion
history, in order to tell one model from the other, it will
suffice to see whether the experimentally determined growth
function lies above or below the theoretical growth function of
the $\Lambda$CDM model. If it lies below, the interacting model
gets automatically discarded; if it lies above it is the
noninteracting one that should be disposed of. Altogether, sooner
or later observational data will constrain $\, f_{g}(z) \,$
sufficiently enough so that without explicit calculating the said
function for any of the two kind of models it will possible  to
discriminate one from the other with full certainty.

\bigskip

\noindent Likewise, the maximum of the matter power spectrum of
interacting models will be shifted to larger scales with respect
noninteracting models. Indeed, the redshift of matter-radiation
equality $z_{eq}$ determines when subhorizon density perturbations
start to grow. A model with a larger fraction of DM (i.e., a
noninteracting one) will reach this equality earlier than a model
with a smaller fraction of DM (an interacting model with $Q > 0$).
The power spectrum of matter density perturbations has its maximum
at $\, k =2 \pi/d_{H(z_{eq})}$, where $d_H(z)$ is the radius of
the horizon at redshift $z$. In any model, when $\, \Omega_{m} \,
+ \, \Omega_{b} = \Omega_{rad}\, (1+z_{eq})$ the perturbations
start to grow. Because of the Meszaros effect \cite{Meszaros}, the
lower the fraction of DM, the later the growth of density
perturbations will commence, making  the models differ  further.
Eventually, data coming from large-scale structure observations
will render this distinction feasible.

\section{Some proposed interaction terms}
\noindent As mentioned in the Introduction, several interaction
terms have been proposed in the literature over the years, all of
them \textemdash due to of our lack of knowledge of the dark
sector at the fundamental level\textemdash  $\;$ were based just
on heuristic arguments and  mathematical simplicity.
\\  \

\noindent Here we will focus on two sets on interaction terms. The
first set is \cite{bolotin,sanchez},
\begin{equation}
Q = \xi H \, \rho_{m}, \quad Q = \xi H \,  \rho_{x},  \quad Q =
\xi H \, (\rho_{m} + \rho_{x}), \quad  Q = \xi H \, \frac{\rho_{m}
\, \rho_{x}}{\rho_{m} + \rho_{x}}, \quad Q = - \xi \, (\rho_{m} \,
+ \, \rho_{x})^{.}, \label{eq:Q1-5}
\end{equation}
where $\, \xi \, $, stands for a constant and small, $\mid \xi
\mid \ll 1$, dimensionless parameter.
\\  \

\noindent A second set,  given by \cite{prd_roy}
\begin{equation}
Q = \Gamma \rho_{m}\, , \qquad Q = \Gamma \rho_{x}\, , \quad {\rm
and} \quad  Q = \Gamma (\rho_{m} + \rho_{x}) \, , \label{eq:Q6}
\end{equation}
where the constant $\Gamma$ has the dimensions of inverse of time
and was inspired by proposals about the curvaton decay
\cite{prd_malik} and the decay  of DM into relativistic particles
\cite{apjl_cen}.
\\  \
\bigskip

\noindent By equating the right hand-sides of (\ref{eq:Q1a})  and
(\ref{eq:Q1-5}.1) and integrating the resulting expression, with
$w =$ constant, one obtains
\begin{equation}
r = \frac{(1+c)\, r_{0} \, a^{\xi(1+c)}}{1+c+r_{0} \, - \, r_{0}
\, a^{\xi(1+c)}} \, , \label{eq:r(a)Q1}
\end{equation}
where $c \equiv 3w/\xi$.
\begin{figure}[!htb]
  \begin{center}
    \begin{tabular}{c}
      \resizebox{100mm}{!}{\includegraphics{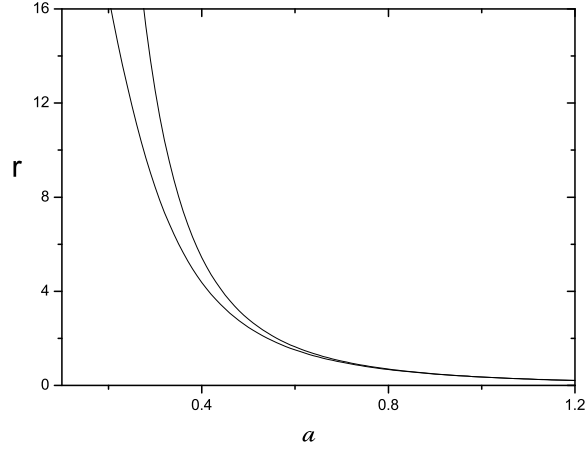}}\\
    \end{tabular}
    \caption{{\small Evolution of $\rho_{m}/\rho_{x}$ for $\, \xi = 0.1 \, $ and $\, \xi =0.01$,
    according to Eq. (\ref{eq:r(a)Q1}). In drawing the curves we have assumed $w = -1$ and
    $r_{0} = 25/70$. These overlap for $a \geq 0.75$.}}
    \label{fig:r(a)Q1}
 \end{center}
\end{figure}
\\  \

\noindent As it can be readily checked, for $0 < \xi \ll 1 \, $,
$\, r \, $ is non-negative for any value of the scale factor, see
Fig. \ref{fig:r(a)Q1}. Further, it does not diverge when $\, a
\rightarrow 0 \, $; in fact $r_{+} = -(1+c)>0$. This is apparent
in the left panel of Fig.1 of Ferreira {\it et al.}
  \cite{prd-pedrof}.
\\  \
\noindent For $\xi <0$ satisfying $\mid \xi \mid \ll 1$, $r \, $
takes negative values and diverges when $a = a_{*}$ with
\begin{equation}
a_{*}= \left(\frac{r_{0}}{1+c+r_{0}}\right)^{\mid \xi \mid (1+c)}
< 1 \, . \label{eq:diver1}
\end{equation}
\\  \
\bigskip

\noindent Proceeding along parallel lines with the second
interaction term, Eq. (\ref{eq:Q1-5}.2), one follows
\begin{equation}
r = \frac{[(1+c)r_{0}\, + \, 1]\, a^{\xi(1+c)}\, - \, 1}{1+c}\,.
\label{eq:r(a)Q2}
\end{equation}
\begin{figure}[!htb]
  \begin{center}
    \begin{tabular}{c}
      \resizebox{100mm}{!}{\includegraphics{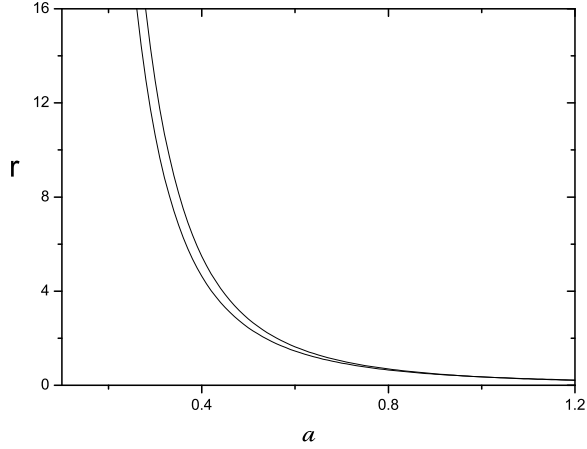}}\\
    \end{tabular}
    \caption{{\small The same as Fig. \ref{fig:r(a)Q1} but for Eq.
   (\ref {eq:r(a)Q2}).}}
    \label{fig:r(a)Q2}
 \end{center}
\end{figure}
\noindent As in the previous case, for $0 < \xi \ll 1$, $\, r$ is
non-negative for any value of the scale factor, see Fig.
\ref{fig:r(a)Q2}. By contrast, for $\xi <0$, with $\mid \xi \mid
\ll 1$, it becomes negative for $\, a > a_{*} = [1\, + \,
r_{0}(1+c)]^{\mid \xi \mid (1+c)} > 1$. Further, regardless of the
sign of $\xi$, $\, r_{+}\,$ diverges (see middle panel of Fig. 1
in \cite{prd-pedrof}).
\\  \
\bigskip

\begin{figure}[!htb]
  \begin{center}
    \begin{tabular}{c}
      \resizebox{100mm}{!}{\includegraphics{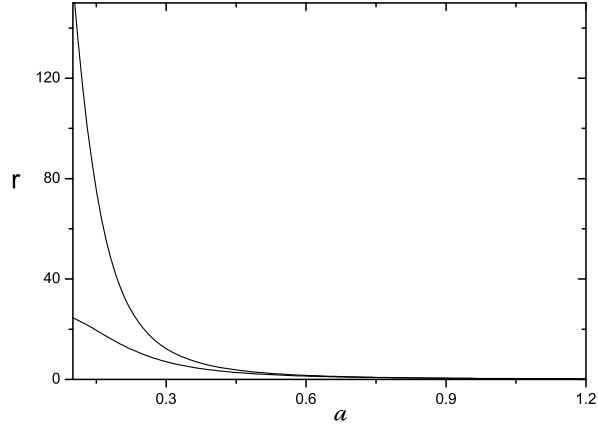}}\\
    \end{tabular}
    \caption{{\small Two numerical solutions to (\ref{eq:dQ3}). The lower and upper curves
    correspond to  $\, \xi = 0.1\, $ and $\, \xi = 0.01\,$, respectively.
    In drawing the curves we have taken  $\, r_{0} = 25/70 \,$ and $\, w = -1$.}}
    \label{fig:r(a)Q3}
 \end{center}
\end{figure}

\noindent From Eqs. (\ref{eq:Q1a}) and (\ref{eq:Q1-5}.3), one is
led to
\begin{equation}
\frac{{\rm d}r}{r^{2}\, + \, (2+c)r \, + \, 1} = \xi \, \frac{{\rm
d}a}{a} \, . \label{eq:dQ3}
\end{equation}
Two numerical solutions to this equation are depicted in Fig.
\ref{fig:r(a)Q3}.
Though not shown, $r$ remains finite in the $a \rightarrow 0$
limit. In fact, $r_{+} =(2/3\xi)-1+\sqrt{(9/4\xi^{2})-(3/\xi)}$,
as is apparent in the right panel of Fig. 1 of \cite{prd-pedrof}
and in Fig. 2 of Olivares {\it et al.} \cite{prd-german}.
\\  \
\bigskip

\begin{figure}[!htb]
  \begin{center}
    \begin{tabular}{c}
      \resizebox{100mm}{!}{\includegraphics{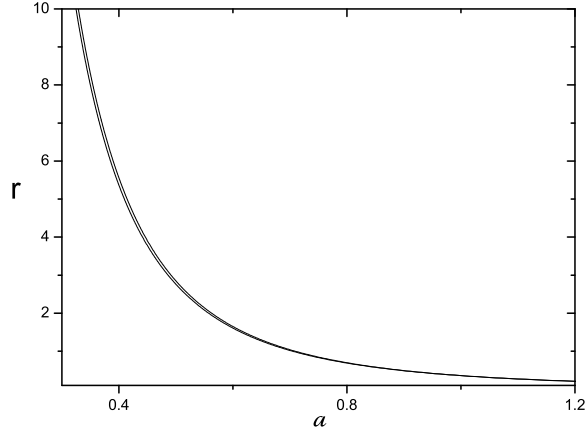}}\\
    \end{tabular}
    \caption{{\small Evolution of the ratio $\, \rho_{m}/\rho_{x}$ vs the scale factor,
    according to (\ref{eq:r(a)Q4}),  for $\, \xi = 0.1\, $ and $\, \xi = 0.01\,$.
    In drawing the curves we have assumed $\, r_{0} = 25/70 \,$ and  $\, w = -1$.}}
    \label{fig:r(a)Q4}
 \end{center}
\end{figure}

\noindent For the interaction term given by  Eq. (\ref{eq:Q1-5}.4)
one obtains
\begin{equation}
r\, \left(r \, + \, \frac{1+c}{c}\right)^{1/c} = r_{0}\,
\left(r_{0} \, + \, \frac{1+c}{c}\right)^{1/c}\, a^{\xi (1+c)} \,
. \label{eq:r(a)Q4}
\end{equation}
Figure \ref{fig:r(a)Q4} shows the behavior of $ \, r \,$ for $\,
\xi = 0.1\, $ and $\, \xi = 0.01$. Both graphs essentially
overlap; on the other hand, there would have been practically no
difference if we would have used $\xi = - 0.1\, $ and $\, \xi = -
0.01\, $ instead. In this case $\, r_{+}\,$ diverges for any sign
of $\xi$. This is consistent with the fact that (\ref{eq:Q1-5}.4)
reduces to (\ref{eq:Q1-5}.2) when $\rho_{m} \gg \rho_{x}$; i.e.,
when $a \rightarrow 0$.
\\  \
\bigskip

\noindent Let us consider now the interaction term
(\ref{eq:Q1-5}.5). With the help of the conservation equations (2)
and (3), it can be cast as
\begin{equation}
Q = 3 \xi H \rho_{m}\, \frac{1+r+w}{r} \, .
\label{eq:Q5a}
\end{equation}
Following the same steps as before, we find
\begin{equation}
 \frac{{\rm d}r}{\xi\, (1+r)^{2} \, + \, wr\, (1+\xi)\, + \, \xi w} = 3 \, \frac{{\rm d}a}{a}
 \, .
\label{eq:dQ5}
\end{equation}
This equation integrates to
\begin{equation}
a = \exp \left[F(r) \, - F(r_{0}) \right] \, , \label{eq:a(r)Q5}
\end{equation}
\begin{figure}[!htb]
  \begin{center}
    \begin{tabular}{c}
      \resizebox{100mm}{!}{\includegraphics{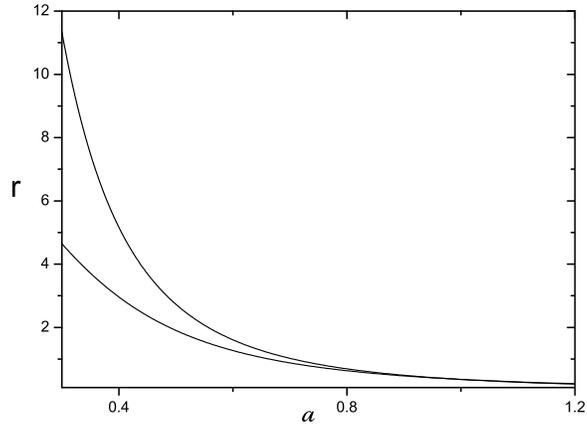}}\\
    \end{tabular}
    \caption{{\small Ratio $\, \rho_{m}/\rho_{x}$ vs the scale factor
    according to Eq. (\ref{eq:a(r)Q5}). The upper and lower
    curves correspond to $\, \xi = 0.1 \, $ and $\, \xi = 0.01$.
    In drawing the curves we have used $\, r_{0} = 25/70 \,$ and $\, w = -1$.}}
    \label{fig:r(a)Q5}
 \end{center}
\end{figure}
where
\begin{equation}
F(r) = \frac{2}{3\sqrt{\mid w\mid} \, \sqrt{(1+\xi)^{2}\mid w\mid
\, - \, 4\xi}} \, \tanh^{-1} \left[ \frac{(1+\xi)\mid w\mid\, - \,
2\xi(1+r)}{\sqrt{\mid w\mid} \, \sqrt{(1+\xi)^{2}\mid w\mid \, -
\, 4\xi}}\right] \, . \label{eq:F(r)}
\end{equation}
\\  \

\noindent Figure \ref{fig:r(a)Q5} shows the evolution of $r(a)$
for $\xi = 0.1 \, $ and $\, \xi = 0.01$. In this case, $r_{+}$
remains finite, and for  $w = -1$, it reads
\begin{equation}
r_{+} = \frac{(1-\xi)\, + \, \sqrt{(1+\xi)^{2}\, - \, 4 \xi}}{2
\xi}.
\label{eq:5r+}
\end{equation}
Obviously, $\xi \, $ cannot  be negative.
\\  \

\noindent Altogether, the interactions terms  given by Eqs.
(\ref{eq:Q1-5}.1), (\ref{eq:Q1-5}.3), and (\ref{eq:Q1-5}.5)
fulfill the criteria of Sec. II provided $\xi > 0$; not so the
other two since $\, r_{+} \, $ diverges.
\\  \
\bigskip

\noindent As for the interaction term given by Eq.
(\ref{eq:Q6}.1), we proceed as follows. Let $\, \Gamma = \gamma
H_{0}$, where $\gamma \, $ denotes a dimensionless constant. Next,
we equate the right-hand side of the resulting expression for $Q$
with the corresponding one of (\ref{eq:Q1a}). After simplifying
and introducing the ansatz $H = H_{0} \, r^{\alpha}$ (piecewise
valid for $\alpha =$ constant in the range $0 < \alpha < 1$), we
arrive at
\begin{equation}
\frac{{\rm d}r}{3wr \, + \, \gamma \, r^{1-\alpha}(1+r)} =
\frac{{\rm d}a}{a} \, .  \label{eq:dQ6}
\end{equation}
\\  \

\noindent It is seen  that  ${\rm d}r/{\rm d}a$ will be negative
in general only if $\gamma < 0$. However, as the left panel of
Fig. 1 in \cite{prd-pedrof} suggests, $r_{+}$ diverges.
\\  \

\noindent A similar situation occurs  for the interaction terms
(\ref{eq:Q6}.2) and (\ref{eq:Q6}.3). Thus, we infer that the three
expressions  for $Q$ in (\ref{eq:Q6})  fail as physically
acceptable interaction terms in the dark sector.

\section{Concluding remarks}
\noindent Ideally speaking, any valid expression for the
interaction between the DM and DE components could be obtained by
writing the equations of motion after varying the corresponding
Lagrangian. However, one cannot propose such Lagrangian expression
without, implicitly or explicitly, presuming a good deal about the
microscopic nature of these components. Worse than that, as
recently noted, because two fluids should not be considered ideal
on their own if they interact with one another, it is unclear if
Lagrangian (even if is just effective) could be ascribed to such a
situation \cite{valerio}. This is why we have followed a diverse
route when looking for phenomenological, but useful, expressions
of the said interaction term.
\\  \

\noindent Any interaction in the dark sector is bound to change
the ratio between the energy densities of DM and DE with respect
to the very particular case of no interaction,  as it is directly
connected to it  [see Eq. (\ref{eq:Q1a})]. Thus, in Sec. II, by
parametrizing this ratio on simple, sensible grounds, reasonable
expressions for $Q$ follow, as Eqs. (\ref{eq:Q1stparm}),
(\ref{eq:Q2ndparm}), and (\ref{eq:Q3rdparm}) illustrate. Here we
wish to stress that this method  rests solely on the FRW metric
and the energy conservation equations and not on any specific
cosmological model or theory of gravity. (Obviously, other
expressions for $\, Q $ differing from the ones proposed here in
Sec. II are obtainable using other possible parametrizations of
$\, \rho_{m}/\rho_{x}$.) Likewise, we have restricted ourselves to
the case $\, Q > 0 \, $ because this helps alleviate the
coincidence problem.
\\  \

\noindent Useful constraints on the free parameters have been
readily obtained (Sec. III) by imposing the vanishing of the
deceleration parameter at the transition redshift $\, z_{da} =
0.74 \pm 0.05$, as recently determined \cite{farooq2013}.
Admittedly, the latter implies the use of one or another theory of
gravity \textemdash in the present case, general relativity. At
any rate, the above parametrizations seem to comfortably reproduce
the history of the Hubble factor (see Figs. \ref{fig:farooq1} and
\ref{fig:farooq2}) which suggests that these terms should behave
reasonably well at the background level because in FRW universes,
$H(z)$ is directly connected to the luminosity and angular
distances. Therefore, they are expected to make good fits to
current cosmological data from supernovae type Ia, baryon acoustic
oscillations, the shift of the first acoustic peak of the cosmic
microwave background temperature spectrum, and gas mass fraction
in galaxy clusters. Nevertheless, we do not claim that these
interaction terms should be taken too seriously since they were
proposed mainly on illustrative purposes and, ultimately, any
cosmological model that assumed whatever interaction term must
also fit observations at the perturbation level, such as the
matter power spectrum and the evolution of the matter growth
factor; but this lies beyond the scope of this paper and will be
the subject of a future research.
\\  \

\noindent We wish to stress that the $\, r(a) \,$ expressions
based on the criteria laid down in SEc. II do not necessarily are
to converge to  the $\Lambda$CDM ratio $\, r_{\Lambda} \propto
a^{-3}$ at any limit. This is so because $r_{\Lambda}$ diverges
very rapidly for $a \rightarrow 0$, which contrasts with the
condition that $r(a)$ varies very little or remains constant at
very early times. Nevertheless, by suitably adjusting their free
parameters, they can approximate $\, r_{\Lambda}$ for $\, a
\gtrsim 0.1$ with great accuracy.
\\  \

\noindent As Sec. IV illustrates, the possible degeneracies
between the interaction term and the EOS of DE are only apparent
because they can be readily broken  at the background and
perturbative levels. Specifically, noninteracting models are
unable to reproduce the same history of the Hubble function of a
interacting model (and vice versa).
\\ \

\noindent Finally, for the sake of completeness,  in Sec. V  the
ratio $\rho_{m}/\rho_{x}$ associated to a handful of interaction
terms found in the literature has been analyzed. Some of them
happen to satisfy the reasonable criteria laid down in  Sec. II.

\acknowledgments{We are indebted to Fernando Atrio-Barandela and
Enrique Gazta\~{n}aga for helpful conversations. SdC and RH were
supported by the COMISION NACIONAL DE CIENCIAS Y TECNOLOGIA
through FONDECYT Grant N$^{0}$s 1110230 and 1130628. SdC was
partially supported by PUCV Grant N$^0$ 123.710 and RH by DI-PUCV
N0 123724. D.P. is indebted to the ``Instituto de F\'{\i}sica de
la Pontificia Universidad Cat\'{o}lica de Valpara\'{\i}so", where
part of this work was done, for warm hospitality and financial
support. This research was partially supported by the ``Ministerio
de Econom\'{\i}a y Competividad, Direcci\'{o}n General de
Investigaci\'{o}n Cient\'{\i}fica y T\'{e}cnica", Grant N$_{\rm
o.}$ FIS2012-32099. Sadly, shortly after this paper was completed
Prof. Sergio del Campo, unexpectedly, passed away. RH and DP
dedicate this paper to his Memory.}



\begin{thebibliography}{99}
\bibitem{polyakov1} A.M. Polyakov, Nucl. Phys. B
\underline{797}, 199 (2008).
\bibitem{polyakov2} A.M. Polyakov, Nucl. Phys. B
\underline{834}, 316 (2010).
\bibitem{krotov-polyakov} D. Krotov and A.M. Polyakov, Nucl. Phys. B
\underline{849}, 410 (2011).
\bibitem{jcap_brax-martin} Ph. Brax and J. Martin, J. Cosmol. Astropart. Phys.
11 (2006) 008.
\bibitem{jmartin} J. Martin (private communication).
\bibitem{coincidence} P.J. Steinhardt, in {\em Critical Problems
in Physics}, edited by V.L. Fitch and D.R. Marlow (Princeton
University Press, Princeton, NJ, 1997).
\bibitem{prd_german} Germ\'{a}n Olivares, Fernando
Atrio-Barandela, and Diego Pav\'{o}n, Phys. Rev. D \underline{77},
103520 (2008).
\bibitem{prd_bin-elcio} A.A. Costa, X-D. Xu, B. Wang,
E.G.M. Ferreira, and E. Abdalla, Phys. Rev. D \underline{89},
103531 (2014).
\bibitem{abdalla-ferreira} E. Abdalla, E.G.M. Ferreira, J. Quintin, and
B. Wang,  arXiv:1412.2777.
\bibitem{bolotin} Yu. L. Bolotin, A. Kostenko, O.A. Lemets, and
D.A. Yerokhin, Int. J. Mod. Phys. D \underline{24}, 1530007
(2015).
\bibitem{fernando} Fernando Atrio-Barandela and Diego Pav\'{o}n,
 in {\em Dark Energy \textemdash Current Advances
and Ideas}, edited by J. R. Choi (Research Singpost, Trivandum,
India, 2009).
\bibitem{copeland} C. Skordis, A. Pourtsidou, and E.J. Copeland, Phys. Rev. D
\underline{91}, 083537 (2015).
\bibitem{langlois} J. Gleyzes, D. Langlois, M. Mancarella,
and F. Vernizzi, arXiv:1504.05481.
\bibitem{peebles-ratra} P.J.E. Peebles and B. Ratra, Rev. Mod.
Phys. \underline{75}, 559 (2003).
\bibitem{hagiwara} K. Hagiwara {\it et al.}, Phys. Rev. D.
\underline{66}, 010001 (2002).
\bibitem{prd_bean} R. Bean, S. H. Hansen, and A. Melchiorri, Phys.
Rev. D \underline{64}, 103508 (2001).
\bibitem{jcap_xia-viel} J.-Q. Xia and M. Viel, J. Cosmol. Astropart. Phys. 04 (2009) 002.
\bibitem{farooq2013} O. Farooq and B. Ratra, Astrophys. J.  \underline{766}, L7 (2013).
\bibitem{ages} B. Chaboyer, Phys. Rep. \underline{307}, 23 (1998);
F.D. Grundahl {\it et al.}, Astron. J. \underline{120}, 1884
(2000); R. Caryel {\it et al.}, Nature (London) \underline{409},
691 (2001).
\bibitem{Ade2013} P.A.R. Ade {\it et al.} (Planck Collaboration), Astron. Astrophys.
\underline{571}, A16 (2014).
\bibitem{Riess2011} A.G. Riess {\it et al.}, Astrophys. J.
\underline{730}, 119 (2011).
\bibitem{chen2003} G. Chen, J.R. Got III, and B. Ratra, Publ.
Astron. Soc. Pac. \underline{115}, 1269 (2003).
\bibitem{busca2012} N.G. Busca {\it et al.}, Astron. Astrophys. \underline{552}, A96  (2013).
\bibitem{blake2012} C. Blake, S. Brough, M. Colless, {\it  et al.}, Mon. Not. R. Astron. Soc.
\underline{425}, 405 (2012).
\bibitem{chuang2012b} C.H. Chuang, and Y. Wang,  Mon. Not. R. Astron.
Soc. \underline{435}, 255 (2013).
\bibitem{moresco2012} M. Moresco, A. Cimatti, R. Jimenez {\it et al.},
J. Cosmol. Astropart. Phys. 08 (20012) 006.
\bibitem{simon2005} J. Simon, L. Verde, and R. Jimenez, Phys. Rev. D  \underline{71}, 123001
(2005).
\bibitem{stern2010} D. Stern, R. Jimenez, L. Verde, M.  Kamionkowski, and S.A.  Stanford,
J. Cosmol. Astropart. Phys.  02 (2010) 008.
\bibitem{zhang2012} C. Zhang, H. Zhang, S. Yuan, T.-Jie Zhang, and Y.-C.
Sun, Res. Astron. Astrophys. \underline{14}, 1221 (2014).
\bibitem{prd_serra} P. Serra {\it et al.}, Phys. Rev. D \underline{80},
121302(R) (2009).
\bibitem{euclid} {\sf http://www.euclid-ec.org}.
\bibitem{DiPorto} E. Di Porto, L. Amendola, and E. Branchini, Mon. Not. R. Astron. Soc.
\underline{419}, 985 (2012).
\bibitem{Luca} L. Amendola {\it et al.}, Living Rev. Relativity \underline{16}, 6
(2013).
\bibitem{Meszaros} P. M\'{e}sz\'{a}ros, Astron. Astrophys.
\underline{37}, 225 (1974).
\bibitem{sanchez} I.E. S\'{a}nchez, Gen. Relativ. Gravit. \underline{46}, 1769 (2014).
\bibitem{prd_roy} G. Caldera-Cabral, R. Maartens, and L.A.
Ure\~{n}a-L\'{o}pez, Phys. Rev. D \underline{79}, 063518 (2009).
\bibitem{prd_malik} K. A. Malik, D. Wands, and C. Ungarelli, Phys. Rev. D \underline{67},
063516 (2003).
\bibitem{apjl_cen} R. Cen, Astrophys. J. \underline{546}, L77
(2001).
\bibitem{prd-pedrof} Pedro C. Ferreira, Diego Pav\'{o}n, and Joel C. Carvalho,
Phys. Rev. D \underline{88}, 083503 (2013).
\bibitem{prd-german} Germ\'{a}n Olivares, Fernando Atrio-Barandela,
and Diego Pav\'{o}n, Phys. Rev. D \underline{71}, 063523 (2005).
\bibitem{valerio} V. Faraoni, J.B. Dent, and E.N. Saridakis,
Phys. Rev. D \underline{90}, 063510 (2014).
\end{thebibliography}
\end{document}